\documentclass[usenatbib]{mnras}

\usepackage{newtxtext,newtxmath}

\usepackage[T1]{fontenc}

\DeclareRobustCommand{\VAN}[3]{#2}
\let\VANthebibliography\thebibliography
\def\thebibliography{\DeclareRobustCommand{\VAN}[3]{##3}\VANthebibliography}

\usepackage{float}
\usepackage{stfloats} 

\usepackage{graphicx}	
\usepackage{amsmath}	
\usepackage{gensymb}
\usepackage{xcolor}
\usepackage{soul} 
\definecolor{orange}{rgb}{1,0.5,0}

\newcommand{\msol}{\text{M}_\odot}

\title[Transient Rates from Cosmological Simulations]{Estimating Transient Rates from Cosmological Simulations and BPASS}

\author[M.M. Briel et al.]{
Max M. Briel$^{1}$\thanks{E-mail: max.briel@gmail.com},
J.J. Eldridge$^{1}$,
Elizabeth R. Stanway$^{2}$,
H.F. Stevance$^{1}$,
A.A. Chrimes$^{3}$
\\
$^{1}$Department of Physics, University of Auckland, Private Bag 92019, Auckland, New Zealand\\
$^{2}$Department of Physics, University of Warwick, Gibbet Hill Road, Coventry, CV4 7AL, UK\\
$^{3}$Department of Astrophysics/IMAPP, Radboud University, P.O. Box 9010, 6500 GL, The Netherlands
}

\date{Accepted XXX. Received YYY; in original form ZZZ}

\pubyear{2020}

\begin{document}
\label{firstpage}
\pagerange{\pageref{firstpage}--\pageref{lastpage}}
\maketitle

\begin{abstract}
The detection rate of electromagnetic (EM) and gravitational wave (GW) transients is growing exponentially. As the accuracy of the transient rates will significantly improve over the coming decades, so will our understanding of their evolution through cosmic history. To this end, we present predicted rates for EM and GW transients over the age of the Universe using Binary Population and Spectral Synthesis (BPASS) results combined with four cosmic star formation histories (SFH). These include a widely used empirical SFH of Madau \& Dickinson and those from three cosmological simulations:  MilliMillennium, EAGLE and IllustrisTNG. We find that the choice of SFH significantly changes our predictions: transients with short delay times are most affected by the star formation rate, while long-delay time events tend to depend on the metallicity evolution of star formation. Importantly we find that the cosmological simulations have very different metallicity evolution that cannot be reproduced by the widely used metallicity model of Langer \& Norman, which impacts the binary black hole merger and stripped-envelope supernovae rates in the local Universe most acutely. We recommend against using simple prescriptions for the metallicity evolution of the Universe when predicting the rates of events that can have long delay times and that are sensitive to metallicity evolution.
\end{abstract}




\section{Introduction} \label{sec:intro}

Astrophysical transients are central to our understanding of stellar populations due to their dependence on the evolutionary history of the progenitor stars. Transient studies, therefore, probe distinct evolutionary and environmental parameters, although subject to model uncertainties.
Commonly observed events are electromagnetic transients, which include supernovae (SNe), their many SNe subtypes, Long Gamma-Ray Bursts (LGRB), and Pair-Instability SNe (PISNe), while the gravitational wave (GW) transients comprise black hole-neutron star (BHNS), binary black hole (BBH), and binary neutron star (BNS) mergers, with the latter also producing an electromagnetic signal, know as a short gamma-ray burst \citep{abbott_2017}. 

Predicting cosmic transient rates is an essential test for population synthesis codes and provides insight into the underlying stellar population producing these short-lived astrophysical events. To perform these predictions, environment and stellar evolution prescriptions are required. The influence of the former on transient rates has started to be considered in detail in the last few years and primarily focuses on GW transients \citep{kruckow_2018, giacobbo_2018, santoliquido_2020, tang_2020, dubuisson_2020, chruslinska_2019, chruslinska_2019a, belczynski_2020}. It describes the amount of stellar material formed (cosmic star formation rate density; CSFRD) and metallicity evolution over the history of the Universe. Traditionally, the CSFRD is an empirical fit to volume-averaged UV and IR observations \citep{behroozi_2013, madau_2014, madau_2017} averaging out the effects of cosmic variance due to large scale structures. It is well measured up to a redshift of 8, with limited observations going up to $z=10$. At lower redshifts further constraints on the empirical cosmic star formation history (SFH) can be obtained, such as using radio \citep{karim_2011,matthews_2021}, 24 micron \citep{rodighiero_2010} or, most commonly, H$\alpha$ \citep{gilbank_2010} observations, which find similar results to UV and IR estimates.

The second component of the environment prescriptions (the metallicity) consists of an average cosmic metallicity evolution, a mass function, and mass-metallicity correlations, which are combined to give a fractional mass density per metallicity over redshift, as described in \citet{langer_2006}. Each component is empirically fitted using different techniques from emission-line studies to deep sky surveys. Moreover, it is widely used in population synthesis studies of GW transients, and modified and tested against GW transient rates by \citet{neijssel_2019}.

The combination of CSFRD and metallicity evolution provides a metallicity-specific cosmic SFH that conforms to observations, but it misses the details of at the scale of individual galaxies and is subject to observational biases and sparse data. This additional level of detail can be provided by cosmological simulations, which produce star formation histories and metallicity evolution on a per galaxy basis, though subject to the assumed input physics and numerical limits. As a result, cosmological simulations have been extensively used to look at the relations between host galaxies and GW transients \citep{artale_2019, artale_2020, mapelli_2017,mapelli_2018,mapelli_2019, toffano_2019}. And with improvements to the input physics and output resolution in recent years, cosmological simulations reproduce many observational relations, such as the CSFRD and galaxy colour bimodality at low redshift \citep{nelson_2018}.

The stellar prescription for predicting cosmic transients is encompassed in the physics of a population synthesis code and needs to cover a broad mass range and binary configurations due to the variety in transient progenitors. The influence of the implemented evolutionary physics on the compact object merger rate  has been explored using rapid population synthesis codes \citep[e.g.][]{dominik_2013,mapelli_2017,mapelli_2018,mapelli_2019,santoliquido_2020,artale_2020,broekgaarden_2021}. Such codes allow for the exploration of a vast evolutionary parameter space using a variety of models and analytical fits, but therefore is limited in the inclusion of computationally intensive detailed stellar structures, which can alter the outcome of binary interactions \citep{gallegos-garcia_2021}.

We use the detailed models from Binary Population And Spectral Synthesis (BPASS) \citep{eldridge_2017, stanway_2018a} to predict EM and GW transient rates from a single population. Similar to \citet{eldridge_2019}, we combine these rates with a star formation history to extract the cosmic event rate density. But instead of using an empirical CSFRD and metallicity description, we extract detailed star formation histories (SFHs) and metallicity evolution from the EAGLE \citep{schaye_2015,crain_2015}, IllustrisTNG \citep{springel_2018,nelson_2018,pillepich_2018a,naiman_2018,marinacci_2018}, and MilliMillennium \citep{springel_2005} simulations. By predicting all transients from these different synthetic stellar populations, we put the most robust and self-consistent constraints to date on the population synthesis physics and the star formation environment. Moreover, it removes possible degeneracies that can originate when looking at a single transient rate.

Section \ref{sec:bpass} discusses the population synthesis of BPASS and EM and GW transients it predicts. The empirical and cosmological simulation star formation histories discussed in section \ref{sec:star_formation_environment} and are combined in section \ref{sec:numericalmethod} with the BPASS predictions to extract cosmic transient rates. We discuss the volume-averaged transient rates and compare them against observations in section \ref{sec:results}. Section \ref{sec:discussion} puts the work in the context of other results, combines the rates, and section \ref{sec:conclusion} concludes with a brief summary of our results.


\section{Binary Population and Spectral Synthesis} \label{sec:bpass}

The BPASS comprises of a grid of 1D theoretical stellar models with single and binary star evolution, which includes stellar winds, mass transfer through Roche Lobe overflow, common envelope evolution, rejuvenation, and chemically homogeneous evolution. The BPASS v2.2.1 models \citep{eldridge_2017, stanway_2018a}, used in this paper, are weighted using a \citet{kroupa_2001} initial mass function extended to $300\, \msol$ and initial binary parameters from empirical distributions \citep{moe_2017}, creating a population with known mass and age. The grid of stellar models contains 13 metallicities ranging from $Z=10^{-5}$ to $0.04$ with $Z=0.020$ considered Solar metallicity.

A vital benefit of the detailed stellar models is the access to the star's structure at the end of its life, which provides a systematic method to predict the subsequent transient. An extensive definition of the transient definitions in BPASS can be found in \citet{eldridge_2013, eldridge_2017}, here we provide a short overview of the criteria.

Massive stars ($\text{M} \gtrsim 8\, \msol$) end their life in an explosion caused by the collapse of the stellar core, appropriately named a Core-Collapse Supernova (CCSN) \citep[e.g.][]{heger_2003, smartt_2009}.
BPASS classifies CCSN progenitors as stars that have a carbon-oxygen core above $1.38\, \msol$, and have a final stellar mass above $1.5\, \msol$. These models have undergone carbon burning and generate a oxygen-neon core capable of collapsing resulting in a supernova \citep{tout_2011}.

From an observational point of view, CCSNe are divided into subtypes based on their spectra; Type II SNe are hydrogen-rich, while Types Ib and Ic lack hydrogen completely but exhibit either the presence or absence, respectively, of helium \citep{filippenko_1997}. Within BPASS, the elemental abundances in the ejecta mass are a proxy to determine the subtype of the CCSN \citep{eldridge_2011,eldridge_2013,eldridge_2017}. The parameters were chosen for  each CCSN subtype to match the observed local relative rates within ${\sim} 20$ Mpc \citep{eldridge_2013} and the Ib to Ic relative rate in \citet{shivvers_2017}. Multiple formation pathways contribute to each of the progenitors; for example, an envelope can be removed by strong stellar winds \citep{heger_2003, woosley_2002}, stable mass transfer with a binary companion, or the explosion of a companion \citep{paczynski_1971, podsiadlowski_1992b,dedonder_1998a, vanbeveren_1998, smith_2011}. The CCSNe progenitors that have lost most or all of their hydrogen envelope are also known as stripped-envelope SNe (SESNe) and are likely to originate from binary systems \citep{yoon_2010, yoon_2015}.  During some SESNe, a relativistic jet is launched, and its interaction with the stellar envelope or the surrounding medium can excite high-energy emission, detectable as a long Gamma-Ray Burst (LGRB) \citep[e.g.][]{heger_2003, langer_2012}. BPASS only considers LGRB formation through chemically homogeneous evolution through fast rotation and mixing caused by low metallicity mass transfer. If the star's evolution leads to a CCSN with a remnant mass greater than $3\, \msol$, it is classified as an LGRB in BPASS. Tidal and magnetar-induced GRBs are added through post-processing by models from \citet{chrimes_2020}.

At the end of their main-sequence evolution, extremely massive stars ($\text{M} \gtrsim 100-130\,\msol$) have a low density, high-temperature helium core. This combination gives rise to electron-positron pair production and removes the radiative pressure that keeps the star from collapsing. The subsequent pair-instability SN (PISN) completely disrupts the star, and no remnant is left behind \citep{fowler_1964, rakavy_1967, heger_2002}. While the exact mass range is uncertain \citep[][and references therein]{farmer_2019, woosley_2021}, BPASS defines a PISN progenitor as a star with a helium core mass between 64 and 133$\, \msol$. Observationally no confident detection has been made, but several Super-Luminous Type I SNe have been identified as possible candidates \citep{woosley_2007, cooke_2012, gal-yam_2009,terreran_2017,gomez_2019}.

Like some SESNe, Type Ia SNe also lack hydrogen in their spectra but are not the result of a core-collapse. Instead, they originate from binary progenitors that involve an electron-degenerate white dwarf. This star gains mass from a main-sequence companion (single-degenerate channel) or merges with another white dwarf through the emission of gravitational waves (double-degenerate channel). If, as a result of the added mass, the white dwarf approaches or exceeds the Chandrasekhar limit of $1.4 \msol$, a thermonuclear explosion occurs \citep[e.g.][]{howell_2011, maoz_2014}. 

Besides being the main driving force behind the Type Ia double-degenerate formation pathway, the loss of orbital energy through gravitational wave emission also drives the decrease in orbital separation and merger of other double compact object systems, such as binary neutron star (BNS), black hole-neutron star (BHNS), and binary black hole (BBH) systems. Seconds before the merger, an increase in gravitational wave frequency and amplitude allows for measurement of this chirp event using the LIGO/VIRGO detector network \citep{abbott_2016}.  
If a BPASS binary system is still bound after both stars have undergone a supernova explosion and two compact objects have been created, the time until the merger is calculated using the gravitational wave radiation orbital energy loss prescription from  \citet{peters_1964}. The carbon-oxygen core mass at the time of the SN determines the formation of a BH or NS.

\begin{figure*}
    \centering
    \includegraphics[width=\linewidth]{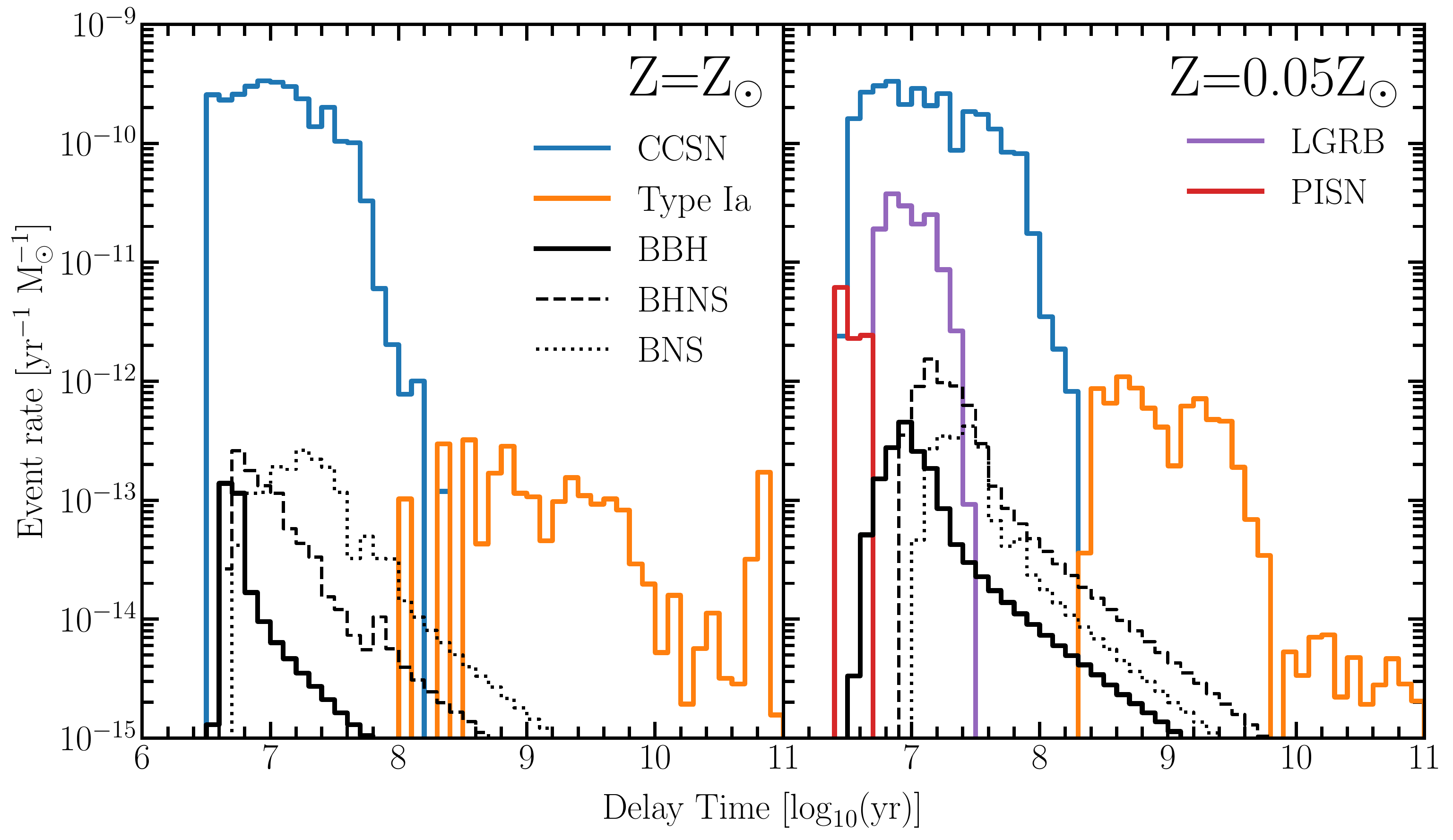}
    \caption{The distribution of time from stellar birth to the specific transient for metallicities $Z=Z_\odot$ (left) and $Z=0.05Z_\odot$ (right).}
    \label{fig:DTD}
\end{figure*}

For the prediction of the cosmic transient rates, we require the number of transients and the time from stellar birth to the transient for each of the BPASS stellar populations, which are 13 delay time distributions (DTDs), each at a different metallicity of which two are shown in Figure \ref{fig:DTD} to show their dependence on metallicity. For example, the distribution of Type Ia SNe and compact object mergers differs between $Z=0.020$ and $Z=0.001$, while the PISNe and LGRBs are constraint to low metallicity populations \citep[][and references therein]{heger_2003}. But, as Figure \ref{fig:DTD} shows, the total CCSN distribution is mostly unaffected by metallicity. It is, however, influenced by the presence of binaries, which increase the delay time of CCSNe to 250 Myr compared to the tens of Myr for single star evolution \citep{zapartas_2017}. These delay time distributions will be combined with the SFH to predict the transient rates. 

\section{Star formation environment} \label{sec:star_formation_environment}

\begin{figure*}
    \centering
    \includegraphics[width=\linewidth]{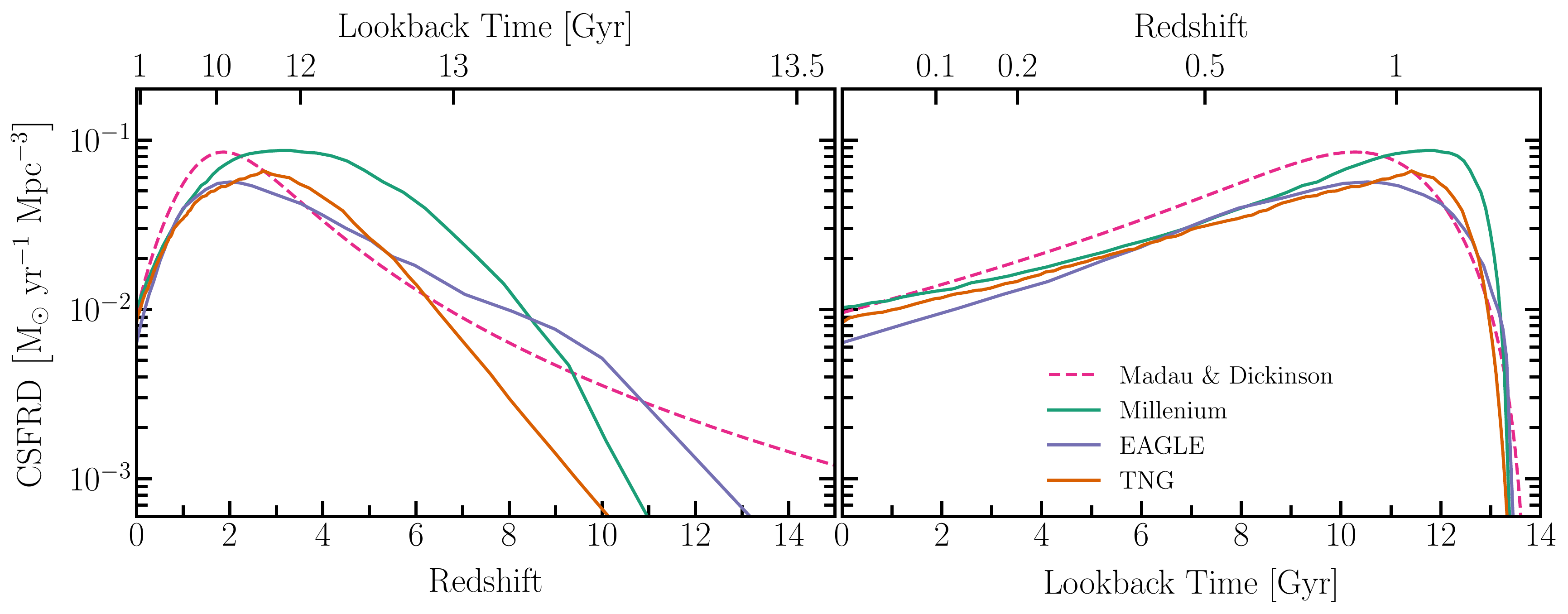}
    \caption{The cosmic star formation rate density over redshift (left) and lookback time (right) for the empirical prescription from Madau \& Dickinson (pink dashed), and rates extracted from the Millennium (green), EAGLE (purple), and IllustrisTNG (orange) simulations.}
    \label{fig:sfh_redshift}
\end{figure*}

\subsection{Semi-Analytical Cosmic Star Formation Rate Density} \label{subsec:semi-analytical}

The CSFRD is fitted from empirically calibrated SFR indicators using UV and IR observations, which are sensitive to the assumed extinction from dust in distant galaxies \citep{wilkins_2016, wilkins_2018}. Early fits include a power-law time-dependence \citep[e.g.][]{behroozi_2013}, but more recent fits typically use a parameterisation for redshift evolution introduced by \citet{madau_2014}:

\begin{equation} \label{eq:M&D}
    \psi(z) = 0.001 \frac{(1+z)^{2.7}}{1 + [(1+z)/2.9]^{5.6}}\ \text{M}_\odot \text{ yr}^{-1} \text{ Mpc}^{-3}
\end{equation}

We transform Equation 15 from \citet{madau_2014} from a Salpeter \citep{salpeter_1955} to a Kroupa IMF by multiplying by 0.66, and this function is shown in Figure \ref{fig:sfh_redshift} as the pink dashed line. It peaks at $z=2$ while declining in the high and low redshift direction, with the current SFR being similar to $z\approx6$. The CSFRD from \citet{madau_2014} is widely used, as such it is our basis for comparison against simulations.

We combine the CSFRD with cosmic metallicity distribution and evolution from \citet{langer_2006} as used in \citet{eldridge_2019} and \citet{tang_2020}:

\begin{equation} \label{eq:langer}
    \centering
    \Psi \left(\frac{Z}{Z_\odot}\right) = \frac{\hat{\Gamma}\left[0.84, (Z/Z_\odot)^2 10^{0.30z}\right]}{\Gamma(0.84)}
\end{equation}

where $\Gamma$ and $\hat{\Gamma}$ are the complete and incomplete gamma functions, respectively, and $Z$ is the volume-averaged metallicity of newly formed stars at redshift $z$. 

We use the 2018 Planck results as our cosmology throughout this paper ($h=0.6766$, $\Omega_\text{M} = 0.3111$, and $\Omega_\Lambda = 0.6889$). To be able to compare all the final rates, all SFH have to be brought to the this cosmology. The empirical SFH extracted from UV observations has an $h$ dependence coming from an $1/h^2$ for the SFR and a $1/h^3$ from the comoving volume. Since the \citet{madau_2014} CSFRD is given in $\msol \text{yr}^{-1} \text{Mpc}^{-3}$, we transform this by first reintroducing the $h$ dependence followed by applying our cosmology. The SFHs from the cosmological simulations and cosmic event rate observations, on the other hand, have an $h^3$ dependence coming from the comoving volume, but similar to the empirical CSFRD we transform them to our cosmology.

\subsection{Simulations} \label{subsec:simulations}

Observationally derived relations have the advantage of being data-driven but are subject to uncertainties in observational completeness and model-dependence in the calibrations required to recover the SFR or metallicity from the data. By contrast, cosmic volume hydrodynamic simulations represent a Universe in which such effects are controlled, and the properties of stars are known precisely. However, they require a tremendous amount of computational power and are subject to their own uncertainties in the assumed physical interactions or sub-grid prescriptions.
These simulations start with dark matter and baryonic particles distributed through a simulation box according to initial conditions based on cosmic microwave background observations. The boxes are evolved up to the current time using simulated large-scale interactions, semi-analytical models for small-scale influences and in some cases also hydrodynamical gas modelling. The assumed strength of interactions and subgrid physics are tuned to match observations or according to theoretical prescriptions. To compare the impact of the SFH on transient rates, we compare the MilliMillennium, EAGLE, and IllustrisTNG simulations to cover different release years, sizes, and physical models.

\subsubsection{MilliMillennium Simulation}

The MilliMillennium Simulation is a subset of the N-body dark matter Millennium Simulation with $\Lambda$CDM cosmology with parameters $h=0.73$, $\Omega_\text{M}=0.25$, and $\Omega_\Lambda = 0.75$ \citep{springel_2005}. Released in 2005, it contains $270^3$ particles in a $62.5\,h^{-1}$\,Mpc box, with each particle representing $8.6 \times 10^8\,h^{-1}\,\text{M}_\odot$ dark matter with a spacial resolution of $5\,h^{-1}$\,kpc. We will refer to the MilliMillennium as Millennium in this paper. As demonstrated by \citet{stanway_2018}, this box is sufficiently large to recover the volume-averaged properties of the bulk galaxy population, although the full simulation would be required to recover rare systems such as extremely massive large scale structures. Starting at $z=127$, 64 selected time steps, known as snapshots, are stored with their gravitationally bound substructure, subhaloes. These are identified by the SUBFIND algorithm \cite{springel_2001} and used to build a merger tree of subhaloes, which is the basic input for the semi-analytical models of galaxy formation \citep{delucia_2007}. When the gas surface density is higher than a critical value, star formation in a disk takes place and follows the parameterisation by \citet{croton_2006}; bulge star formation, however, only occurs during the merger of subhaloes and follows the \citet{somerville_2001} collisional starburst model, which is only able to reproduce the observed gas fraction as a function of galaxy luminosity. 

\subsubsection{EAGLE Simulation}

Evolution and Assembly of GaLaxies and their Environments (EAGLE) \citep{schaye_2015,crain_2015} is a hybrid N-body and hydrodynamical simulation. It contains dark matter particles with a mass of $9.70 \times 10^6\,\text{M}_\odot$, and baryonic gas particles of $1.81 \times 10^6\,\text{M}_\odot$ in a $100\,\text{Mpc}^3$ box for the fiducial model (L0100N1504). The EAGLE simulation has a $\Omega_\text{M} = 0.307$, $\Omega_\Lambda = 0.693$, and $h = 0.6777$ cosmology and does not provide their output with explicit dependence on the Hubble parameter. Therefore, we use the simulations cosmology to reintroduce the $h^3$ dependence and then apply our cosmology to allow for comparison. While only 29 snapshots were recorded between $z=127$ and $z=0$, $1504^3$ particles were used to study galaxy formation. The stellar formation is resolved using sub-grid physics and depends on the pressure in dense gas, and is tuned to reproduce the observed Kennicutt-Schmidt relation \citep{schmidt_1959}. The simulation includes prescriptions for black hole and supernova feedback mechanisms that return baryons to the intergalactic medium and enrich the environment. A full description can be found in \citet{wiersma_2009}. The feedback mechanisms and star formation rate are calibrated to reproduce the galaxy luminosity function at $z=0.1$, BH-stellar mass relation, and galaxy size \citep{crain_2015}. A merger tree is constructed using the same SUBFIND algorithm as the Millennium simulation, but with slight adjustments and inclusion of baryonic matter in substructure identification \citep{dolag_2009}.

\subsubsection{IllustrisTNG Simulation}

Similar to EAGLE, The Next Generation Illustris simulation (IllustrisTNG) is a hybrid simulation that contains dark matter and baryonic matter. These have particle masses $7.5 \times 10^6 \,\text{M}_\odot$ and $1.4 \times 10^6 \,\text{M}_\odot$, respectively, in the TNG100-1 simulation \citep{springel_2018,nelson_2018,pillepich_2018a,naiman_2018,marinacci_2018}, which we shall refer to as the TNG simulation in this paper. $1820^3$ dark matter and $1820^3$ baryonic particles are included in the simulation volume of $110.7^3$ cMpc and are evolved from $z=127$ to present day in a $h=0.6774$, $\Omega_\text{M}=0.30897$, and $\Omega_\Lambda = 0.6911$ $\Lambda$CDM Universe from \citet{ade_2015}. Again, we scale to our own cosmology.
Like the other two simulations, the IllustrisTNG is a hydrodynamic simulation but also includes magnetic fields and new feedback prescriptions. These and the galaxy formation models are fully described in \citet{weinberger_2017} and \citet{pillepich_2018} and aim to agree with observational constraints, such as the cosmic star formation rate density and the stellar mass content of galaxies at $z=0$. Again, the SUBFIND algorithm finds subhaloes (galaxies), but the IllustrisTNG introduces a "SubhaloFlag" to identify gravitationally-bound clusters that are numerical artefacts and not of cosmological origin \citep{nelson_2019}.  We remove non-cosmological subhaloes from our sample. By tracing the baryonic content of each galaxy, the SUBLINK algorithm generates merger trees \citep{rodriguez-gomez_2015}.

\section{Numerical Method} \label{sec:numericalmethod}

\begin{figure*}
    \centering
    \includegraphics[width=\linewidth]{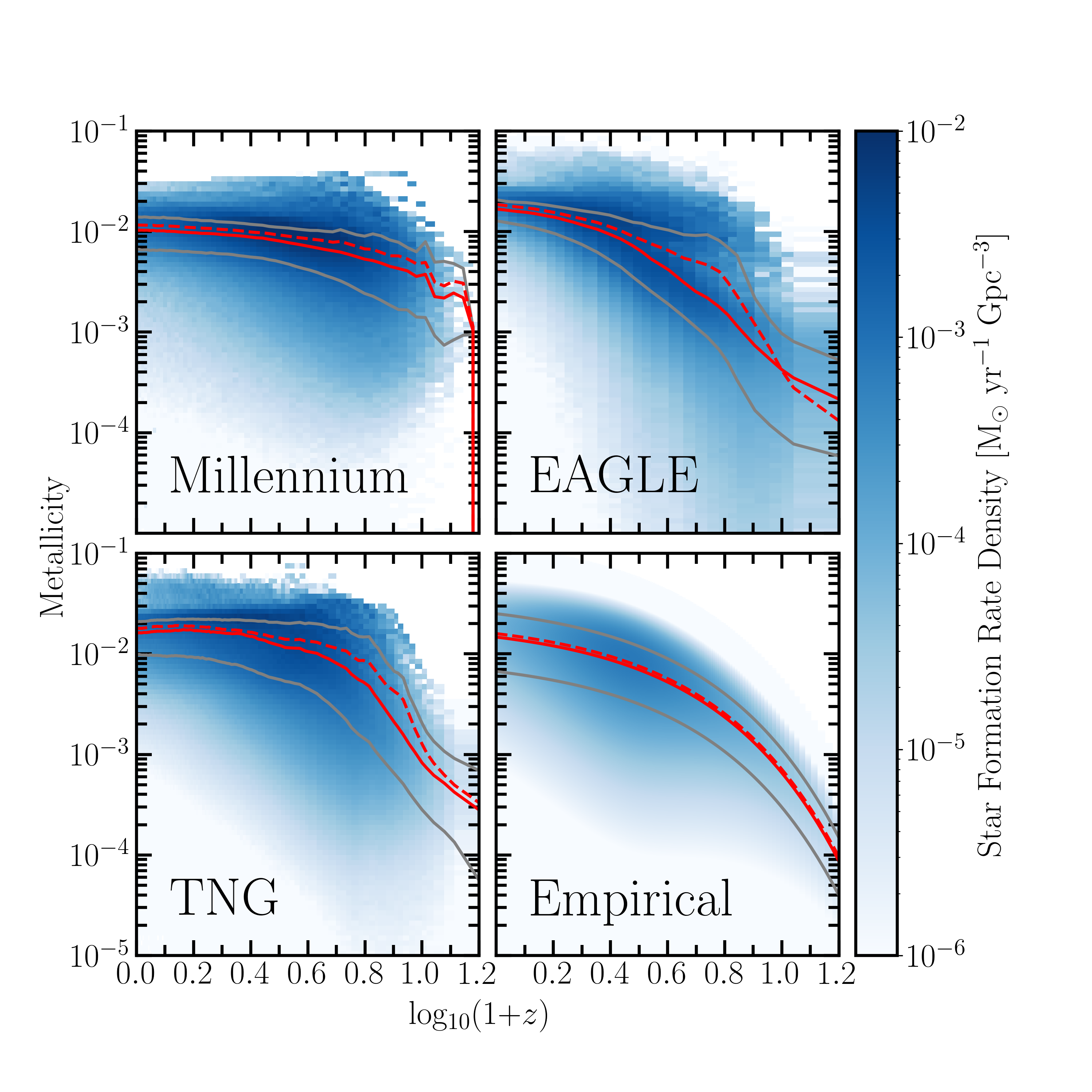}
    \caption{The star formation rate density distribution over metallicity and over $\log_{10}(1+z)$. The median SFR weighted metallicity is indicated with the red solid line with the $1\sigma$ spread shown by the solid grey lines. The dashed red line indicates the SFR weighted mean metallicity. The empirical prescription has a slow increase over redshift, while the Millennium simulation has the opposite evolution with a near flat high median metallicity over redshift. The EAGLE and TNG start at low metallicity at high redshift and have a fast increasing metallicity as the Universe evolves to current day.}
    \label{fig:Z_dist}
\end{figure*}

For the cosmological simulations, we extract the merger trees of each non-zero stellar-mass galaxy at $z=0$ using the public APIs [EAGLE \citep{mcalpine_2016}, TNG; \citep{nelson_2018}, Millennium \citep{lemson_2006}]. Each galaxy has its own individual metallicity and SFR evolution influenced by stellar evolution and galaxy interactions, allowing for a broader range of galaxy metallicities than the empirical CSFRD. The SFR distribution over metallicity and redshift in Figure \ref{fig:Z_dist} shows that the cosmological simulations have a wider metallicity spread and faster enrichment of the Universe than the empirical parameterisation. The latter is especially true for the Millennium simulation, which achieves a high mean metallicity at $z=10$ and stays nearly flat throughout cosmic history. This is similar to the mean metallicity evolution of the TNG simulation, which remains mostly flat with only a fast increase between $\log_{10}(1+z)=0.8$ and $\log_{10}(1+z)=1.0$. The EAGLE and empirical distributions, on the other hand, have a gradual increase in their mean metallicity towards current time. In the case of the EAGLE, we even see evidence for a bimodal distribution in the metallicity evolution with a constant high metallicity population at $Z=0.015$ from $\log_{10}(1+z)=0.8$ to current time, and a second lower metallicity population that is present from the start of star formation and slowly increases in metallicity over redshift. This complex behaviour cannot be reproduced in the analytical models typically used. This, for example, allows for low metallicity events to still occur when high metallicity events are more prevalent.

\begin{table}
\begin{center}
\begin{tabular}{c c c c c}
\hline
 & \textbf{Empirical} & \textbf{Millennium} & \textbf{EAGLE} & \textbf{TNG} \\ 
\hline 
 &     \multicolumn{4}{c}{$\log_{10}(\mathcal{R})$ [yr$^{-1}$ Gpc$^{-3}$]}\\
\hline
\textbf{BBH} & 2.01 & 2.01 & 1.70 & 1.63 \\ 
\textbf{BHNS} & 2.38 & 2.36 & 2.03 & 2.04 \\ 
\textbf{BNS} & 2.35 & 2.33 & 2.16 & 2.23 \\ 
\hline
\textbf{Ia} & 4.25 & 4.22 & 4.06 & 4.12 \\
\textbf{LGRB} & 2.29 & 2.04 & 1.13 & 1.55 \\ 
\textbf{PISN} & 0.82 & 0.71 & -0.09 & 0.10 \\ 
\textbf{CCSN} & 4.92 & 4.96 & 4.74 & 4.86 \\ 
\hline
\textbf{II}  & 4.20 & 4.27 & 3.84 & 4.03 \\ 
\textbf{IIP}  & 4.64 & 4.69 & 4.45 & 4.58 \\  
\textbf{Ib} & 4.15 & 4.13 & 4.07 & 4.17 \\ 
\textbf{Ic} & 3.98 & 4.04 & 3.88 & 3.99 \\ 
\hline
&     \multicolumn{4}{c}{$\log_{10}(\textrm{SFR})$ [M$_\odot$ yr$^{-1}$ Mpc$^{-3}$]}\\
\hline
SFR z=0 & -2.02 & -1.99 & -2.20 & -2.08 \\
SFR z=2 & -1.08 & -1.13 & -1.25 & -1.26 \\
\hline

\end{tabular}
\end{center}
\caption{The cosmic event rate predictions, $\log(\mathcal{R})$, at $z=0$ in a events per year per Gpc$^3$ for the different star formation histories. CCSN contain the Type II, IIP, Ib, and Ic supernovae types. The star formation rate at $z=0$ and $z=1$ are shown for comparison against the rates.}
\label{tab:rates}
\end{table}

Using the metallicity and star formation at each snapshot for each galaxy and the dimensions of the simulation, we construct a volume-averaged CSFRD, as shown in Figure \ref{fig:sfh_redshift}. 
At redshifts below $z=2$, the shapes of the CSFRDs are similar but differ in normalisation with the Millennium CSFRD and semi-analytical prescription having the highest SFRs at $z=0$ followed closely by the TNG and EAGLE CSFRD. The $z=0$ and $z=2$ SFR rates are shown in Table \ref{tab:rates}. 
Above $z=2$, the shape of the Millennium CSFRD and the long tail of the empirical CSFRD stand out. It is significantly different than the other CSFRD by peaking later with a broader spread at $z=3.31$. The empirical CSFRD, on the other hand peaks sharply at $z=1.87$ with the tail continuing into high redshift, while the rates of the cosmological simulations at these early times, when we expect low metallicity to dominate, are low to non-existent.

\begin{figure}
    \centering
    \includegraphics[width=\linewidth]{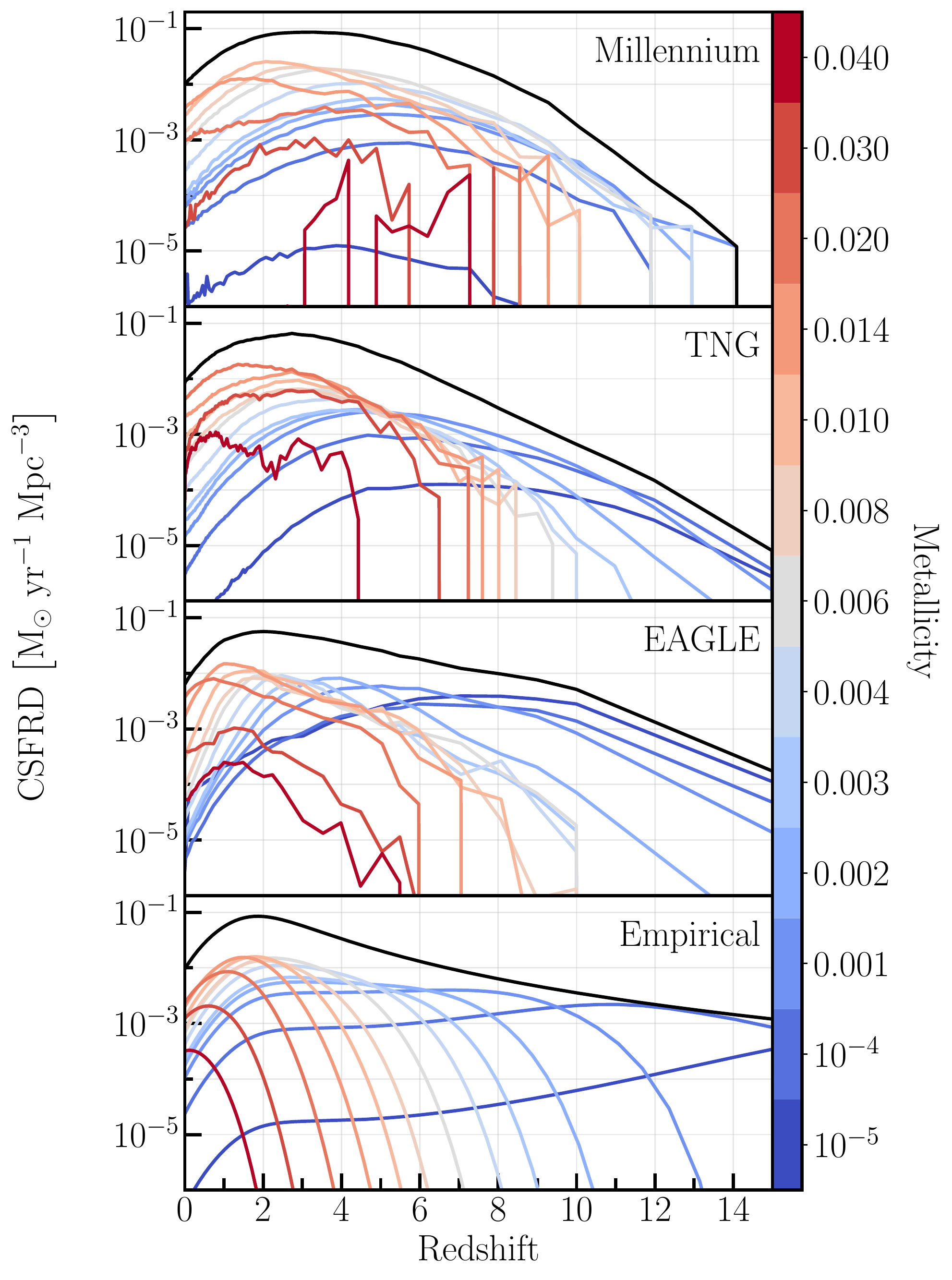}
    \caption{The cosmic star formation rate densities for our empirical prescription, and the Millennium, EAGLE and TNG simulations split into the 13 BPASS metallicity bins indicated by the colour bar. The total star formation rate density is indicated by the black solid line.}
    \label{fig:sfh_metals}
\end{figure}

Based on the mean stellar metallicity reported by the simulation, individual galaxies are binned into one of the 13 BPASS metallicities, resulting in metallicity-specific SFR over redshift, as shown in Figure \ref{fig:sfh_metals}. The empirical CSFRD from \citep{madau_2014} is split into the same metallicity bins using the prescription from \citet{langer_2006}, as described in Section \ref{subsec:semi-analytical}. 
The metallicity distributions in Figure \ref{fig:sfh_metals} indicate an early start in high metallicity star formation in the cosmological simulations. At $z \sim 6$ solar metallicity star formation is ongoing, while the empirical prescription only starts formation at this metallicity at $z=4$. The faster enrichment significantly impacts the rate of specific transients due to their sensitivity to metallicity, see Section \ref{sec:bpass}. Furthermore, a late start in star formation in the cosmological CSFRD reduces the amount of low-metallicity star formation at high redshift.

To estimate transient rates, the metallicity specific SFHs have to be combined with the associated transient DTDs from section \ref{sec:bpass}. This is achieved by splitting the final lookback time into equal-sized bins in linear space. For each bin ($j$) we calculate the rate ($R$) as described in equation \ref{eq:convolution}, where $t_i$ are the bin edges. We integrate the SFH using linear interpolation to account for changes between snapshots. The integrated DTD and SFH are multiplied, summed over each time bin, and the result is divided by the bin width, resulting in the number of events per year per Gpc.

\begin{equation} \label{eq:convolution}
    R_j = \displaystyle\frac{\displaystyle\sum_{i=j+1}^{N} \int_{t_{i-1}}^{t_i} \text{SFH} \int_{t_i - t_{j+1}}^{t_i - t_j} \text{DTD}}{(t_{j+1} - t_{j})}
\end{equation}

The described method has been made available as a module in \textsc{hoki}, which is a python framework and interface for BPASS models \citep{stevance_2020}. The \textsc{csp} module contains several built-in stellar formation histories and allows the user to input their own binned or parameterised SFH to generate complex stellar populations. Moreover, it calculates BPASS transient rates originating from this population at a specific lookback time or over the complete history of the Universe. 

\section{Estimated Transient rates} \label{sec:results}

\subsection{Type Ia}

\begin{figure}
    \centering
    \includegraphics[width=\linewidth]{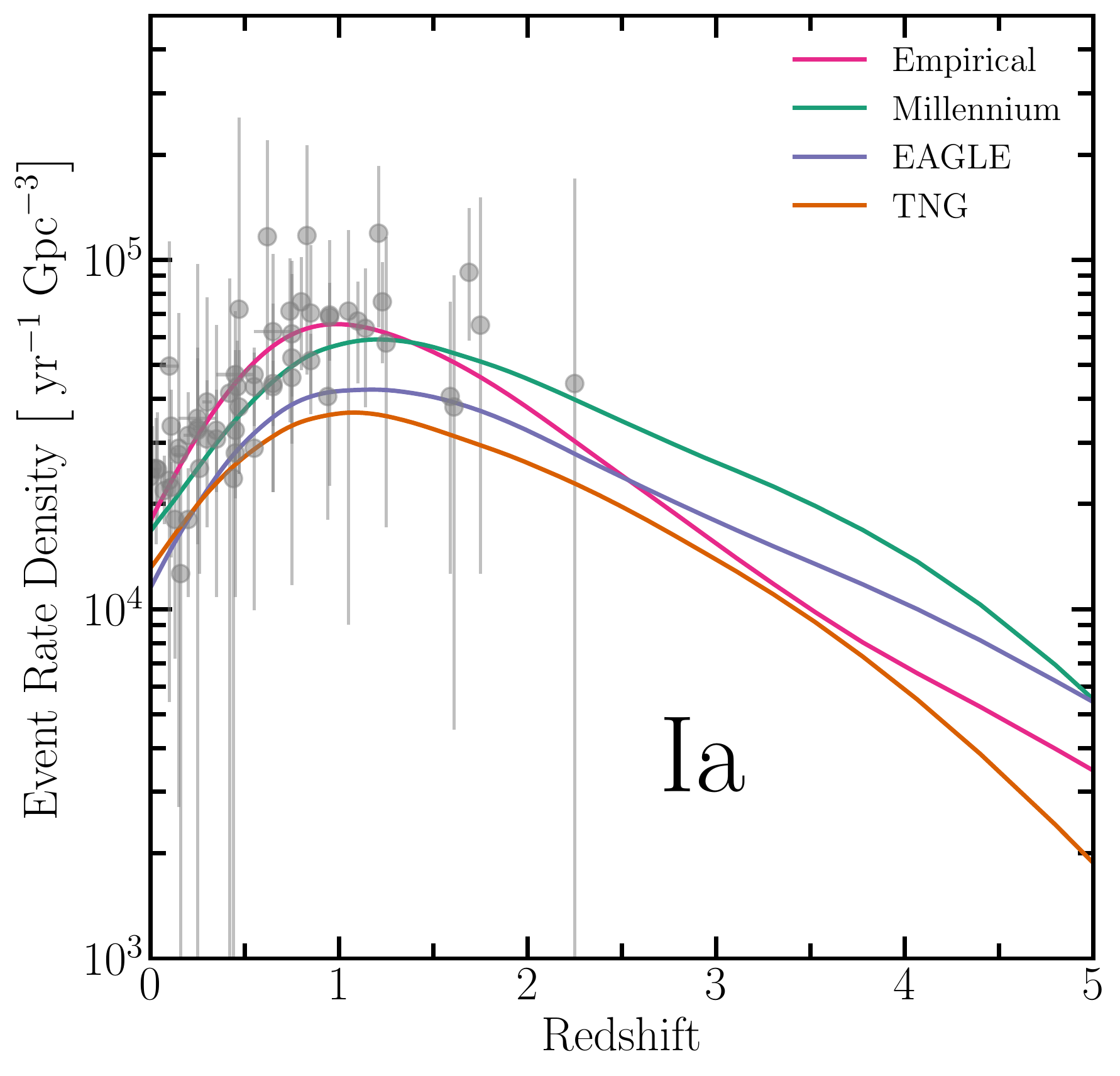}
    \caption{Type Ia SN rate predicted from the four stellar formation histories, compared with observations drawn from a collection of surveys described in Table \ref{tab:Type_Ia_obs}.}
    \label{fig:Type_Ia}
\end{figure}

The formation of a white dwarf and subsequent accretion or merger leading to a thermonuclear explosion takes $10^8$ to $10^{11}$ years depending on metallicity and stellar evolution, due to the inherent binary nature of this event. With the long time between stellar birth and SN, the Type Ia rate probes older star formation \citep{ruiter_2009,mennekens_2010, ruiter_2011, maoz_2012, eldridge_2019}, which we confirm by comparing the peaks of star formation against the peaks of the Type Ia rate in Table \ref{tab:peaks}. The peak Type Ia SN rate occurs $\Delta(z) \approx 1-2$  later than the peak in star formation with the exact delay depending on the simulation and its metallicity evolution.

\begin{table}
    \centering
    \begin{tabular}{c|c|c|c|c|c}
             & \textbf{SFH} & \textbf{Type Ia} & \textbf{CCSN} & \textbf{PISN} & \textbf{BBH} \\
                     & \multicolumn{5}{c}{redshift ($z$)} \\
         \hline
         \textbf{Empirical}  & 1.87 & 0.98 & 1.87 & 2.63 & 1.90\\
         \textbf{Millennium} & 3.31 & 1.21 & 3.31 & 4.40 & 2.78\\
         \textbf{EAGLE}      & 2.01 & 1.17 & 2.01 & 5.87 &2.27\\
         \textbf{TNG}        & 2.73 & 1.09 & 2.73 & 4.40 &2.63\\
    \end{tabular}
    \caption{The redshift ($z$) of the peak in star formation history, Type Ia, CCSN, PISN, and BBH rates.}
    \label{tab:peaks}
\end{table}

The rates over redshift are shown in Figure \ref{fig:Type_Ia} with a collection of observations for comparison. The Millennium and empirical predictions lie around these observations, while the EAGLE and TNG simulations underpredict the rates and only align with a few observed rates at $z<0.5$. The accurate empirical prediction is in contrast to the overestimation in \citet{eldridge_2019} due to reweighing of the CSFRD  from a Salpeter to a Kroupa IMF. The model predictions, however, differ less than a factor 2. To assess the goodness of the predictions, we calculate the reduced $\chi^2$ and see in Table \ref{tab:chi2} that the EAGLE and TNG has reduced $chi^2$ closest to 1. However, we note that due to the large uncertainties of the observations, all reduced $chi^2$ of the Type Ia predictions are below 1, indicating that the choice between these SFH and metallicity evolution are minimally important in predicting the Type Ia rate given current observations. 

\begin{table}
\begin{center}
\begin{tabular}{c|c|c|c|c|c}  
Transient & N & \textbf{Empirical} & \textbf{Millennium} & \textbf{EAGLE} & \textbf{TNG}   \\
\hline
Type Ia  & 60 & 0.32   & 0.28   & 0.81   & 0.95   \\
CCSN     & 25 & 1.25   & 1.01   & 0.62   & 0.58   \\
PISN$^a$ & 5  & 0.38   & 0.84   & 0.61   & 1.17   \\
\hline
\textbf{All} & 90 & 0.58 & 0.51 & 0.74 & 0.86 \\
\hline
LGRB$^{\star}$ & 10 & 1.82  & 3.53  & 4.67 & 7.13   \\ 
\end{tabular}
\end{center}
\caption{The reduced $\chi^2$ value from each model using the given data, where N is the number of observed cosmic event rates for the specific event type. \textbf{All} weights each observed rate equally. $\star$ includes the tidal LGRB contribution from \citet{chrimes_2020} and is, therefore, left out of the \textbf{All}. $^a$: the calculated rate from \citet{zhao_2020} is left out due to the absence of an uncertainty on the observation.
   \label{tab:chi2}}
\end{table}


\subsection{Core-Collapse Supernovae}

\begin{figure}
    \centering
    \includegraphics[width=\linewidth]{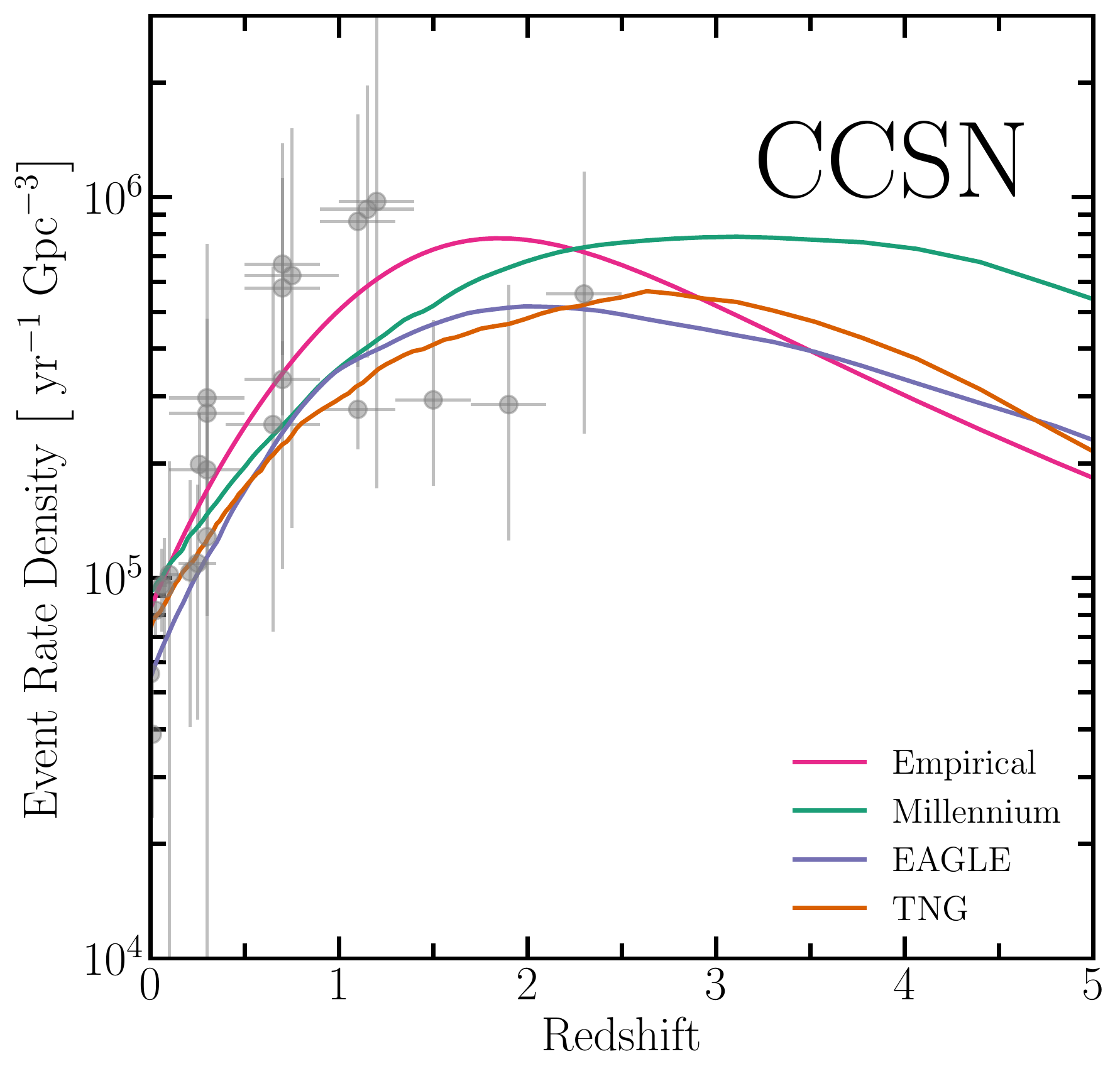}
    \caption{Predicted core-collapse supernova rates with a compilation of observational estimates in grey for comparison (see Table \ref{tab:CCSN_obs}).}   
    \label{fig:CCSN}
\end{figure}

CCSNe occur in young stellar populations, because the CCSN progenitors are massive stars that burn through their nuclear fuel quickly and, thus, have short delay times ($10^{6.5}$- $10^{8.3}$ years). This creates a tight relationship between the SFR and cosmic CCSN rate, as the alignment of the peaks between the CCSN rate and SFH peaks in Table \ref{tab:peaks} shows.
Not only do the peaks align, the CCSN predictions over redshift also closely track the shape of the associated CSFRD, and are shown in Figure \ref{fig:CCSN} with observations. Below $z=0.5$ they follow observations closely till around $z=0.9$ after which the observations are split in two clusters. The first has data points around $z=1$ and $z=1.5$ with high CCSN rates originating from a collection of surveys \citep{melinder_2012,dahlen_2012, graur_2011}. At these redshifts, the empirical CCSN rate prediction has the best fitting rates. The cosmological CCSN rate predictions align better with the second cluster of data points, which are located between $z \sim 1 - 2.5$, have lower rates, and originate from a single survey by \citet{strolger_2015}. Different approaches in correcting for missed SNe due to high extinction could result in this grouping. Because the CCSN rate follows star formation closely, they often happen in very dusty star-forming regions and can be obscured, especially at high redshifts. Hence the fraction of missed SNe has to be accounted for. The higher estimates use a prescription from \citet{mattila_2012}, which can increase the CCSNe rate significantly. Local overdensities of star formation in the survey fields might also cause these observations to be overestimated \citep{dahlen_2012}. \citet{strolger_2015}, on the other hand, use their own method based on the \citet{calzetti_2000} extinction law, but these high redshift observations are challenging and only survey a small area compared to other supernova surveys leaving them more vulnerable to cosmic variance and low number counts. Finally, this grouping and the large standard deviation of the first cluster results in a 1.01 reduced $\chi^2$ for the Millennium simulation and slightly worse values for the empirical, EAGLE and TNG predictions as shown in Table \ref{tab:chi2}.

\subsubsection{CCSN Subtypes}

\begin{figure}
    \centering
    \includegraphics[width=\linewidth]{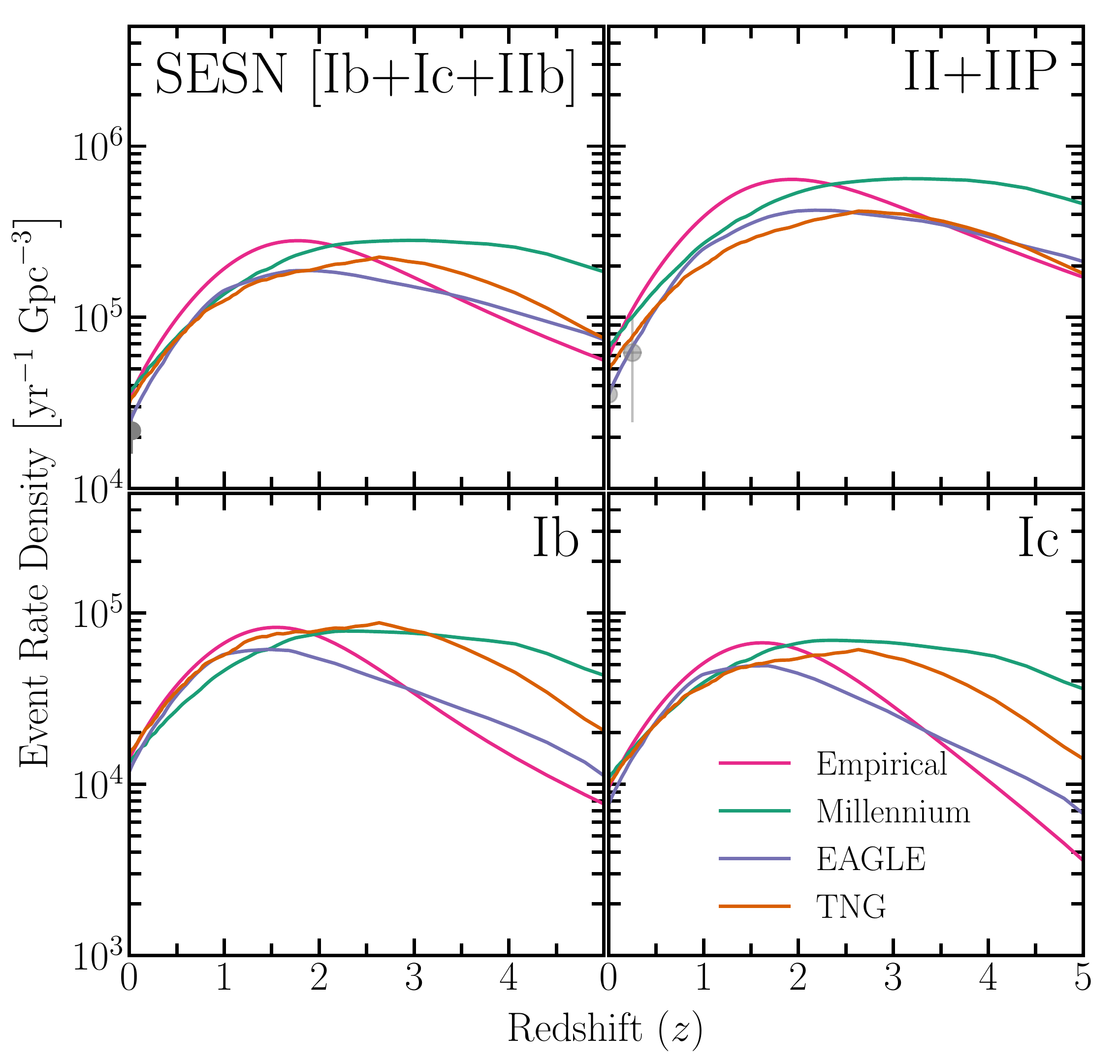}
    \caption{The core-collapse supernova events split up by possible subtypes with a single observed stripped-envelope supernova rate by \citet{frohmaier_2021}, and two Type II observations (See Table \ref{tab:Type_II_obs}).}
    \label{fig:CCSN_subtypes}
\end{figure}

Although the same explosion mechanism causes CCSNe, the subtypes come from progenitors with different mass ranges and mass loss histories, as described in Section \ref{sec:intro}. By looking at the rate of the subtypes and their fraction to the total CCSN rate, we extract more information about the progenitors. However, the limited number of SESN events restricts the calculation of an observed rate. \citet{frohmaier_2021} is one of the few studies to calculate the total combined rate for Type IIb, Ib, and Ic events. Since BPASS does not distinguish between Type II subtypes except for Type IIP events, we use the fraction from \citet{eldridge_2013} of 0.6541 to estimate the Type IIb rate from the non-IIP Type II events. This fraction is not well constrained and differs for other surveys \citep[e.g.][]{smith_2011, li_2011}. Figure \ref{fig:CCSN_subtypes} shows that the predicted rates lie above the \citet{frohmaier_2021} observed rate estimate of $2.18 \times 10^4$~yr$^{-1}$~Gpc$^{-3}$ in our cosmology at $\langle z \rangle = 0.028$. Rates calculated using the EAGLE model approaches this rate with $2.40 \times 10^4$~yr$^{-1}$~Gpc$^{-3}$, but the other SFHs overpredict the observed rate with the Millennium Simulation deviating the most with a rate of $3.65 \times 10^4$~yr$^{-1}$~Gpc$^{-3}$.

The Type II rate predictions align with observed rates from \citet{li_2011a} and \citet{cappellaro_2015}, although the empirical and Millennium SFHs result in an overestimation of this subtype. For the other CCSN subtypes, Type Ib and Type Ic, no observational cosmic rates are available for comparison. Instead, we note that the Type Ib and Ic rates have similar shapes due to similar progenitors and sensitivity to the metallicity evolution. The absolute rate of Type Ib events, however, is slightly higher than Type Ic due to the difficulty of stripping the helium envelope required for a Type Ic SN. Distinguishing between the individual predictions for either of these SN types is impossible at low redshift due to the minimal difference in the rates, which could be attributed to a small difference in metallicity between the predictions. The metallicity distributions in Figure \ref{fig:Z_dist} shows that the EAGLE, TNG and empirical CSFRDs have similar metallicities at these low redshifts, while the Millennium simulation has a slightly lower mean metallicity. However, above a metallicity of half solar, the Ib/c rates in the DTD are near constant, which, together with the similar CSFRD, results in similar rates for the cosmic Type Ib/c predictions. At higher redshifts, the metallicity and CSFRDs become more distinct, which separates the cosmic Type Ib/c rates, as can be seen in Figure \ref{fig:sfh_redshift} and \ref{fig:CCSN_subtypes}.
The Type II SN rates are an order of magnitude higher than the Type Ib and Ic rates and dominate the CCSN predictions, because the progenitor systems do not need to undergo envelope stripping and can be of single or binary star nature, resulting in the predicted rates more closely following the CSFRD.

\subsubsection{CCSN subtype fraction evolution} \label{sec:fractions}

\begin{figure*}
    \centering
    \includegraphics[width=\textwidth]{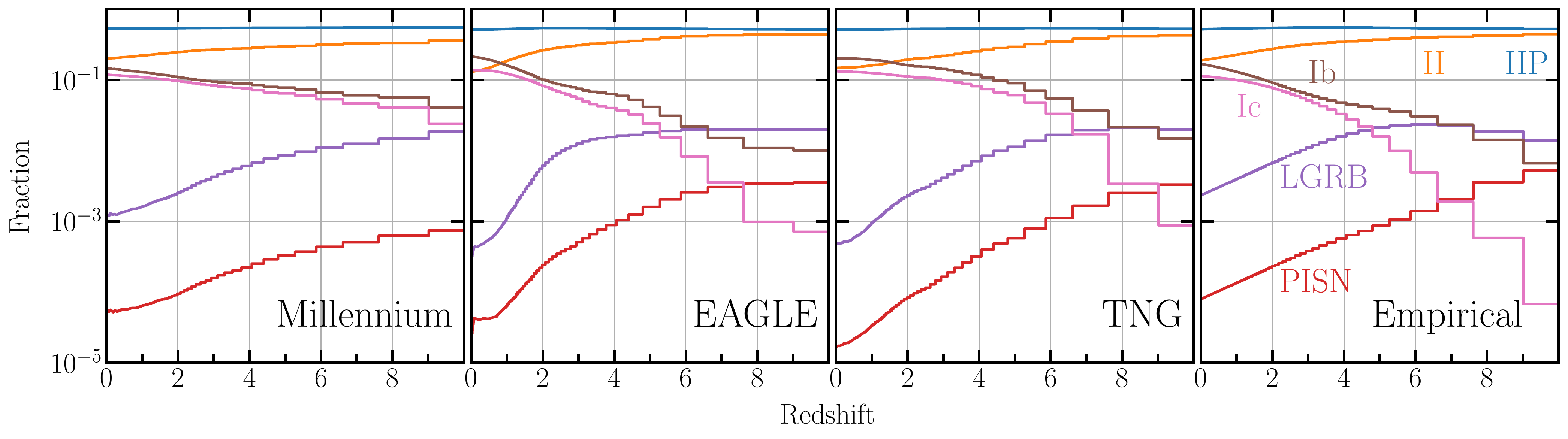}
\caption{The relative fraction of all supernovae that originate from massive stars. LGRB, PISN, and Type II rates (purple, red, orange) decrease as the Universe enriches, while the Type Ib/c (brown and pink) rates increase. The fraction of massive stars ending their life in a Type IIP supernova (blue) is near constant over redshift.}
\label{fig:CCSN_fraction}
\end{figure*}

While the cosmic rates of CCSN subtypes are hard to come by, the fractions of Ib, Ic, and IIb with respect to the total number CCSNe are available from several surveys \citep{li_2011, smith_2011, shivvers_2017, perley_2020}. Therefore, we show our predictions and the fractions from the volume-limited survey up to 60 Mpc from \citet{shivvers_2017} in Table \ref{tab:ccsn_subtype_fractions}. Our Type II predictions are significantly lower than the observed fraction. The magnitude-limited survey from \citet{perley_2020}, however, predicts a lower Type II fraction of 0.722, which is closer to our predictions, although still significantly higher than the 0.64 and 0.67 fractions for the EAGLE and TNG predicted fractions. A slightly higher fraction of 0.766 is found by the 100 Mpc volume-limited survey by  \citet{smith_2020b}.
Instead of the Type II events, there are fractionally more Type Ib/c events in our predictions of which the bulk are Type Ib events. This discrepancy could indicate too swift an enrichment, a too strong stellar wind prescription, a different mass transfer efficiency, or a combination of the above. Since the EAGLE and TNG simulations enrich faster than the Millennium simulation and empirical description, this is the likely origin.

Looking at the evolution of the relative fractions of SNe that originate from massive stars in Figure \ref{fig:CCSN_fraction} makes it clear that the Type IIP remains constant over redshift, because these typically come from single stars and wide binaries. In comparison, the other type II SN fraction decreases with decreasing redshift, being replaced by Type Ib/c. The reason for this is that while binary interactions remove much of the envelope, stellar winds play a significant role in the further evolution. Thus, more metal-rich stellar populations are dominated by SNe that have experienced more mass loss. The scale of this change depends on the metallicity distribution as shown in Figure \ref{fig:Z_dist}, where the mean metallicity evolution of the Millennium simulation is near flat, as is the evolution of the Type Ib/c rate in Figure \ref{fig:CCSN_fraction}. The other SFHs, on the other hand, have a clear metallicity evolution, which shows in the relative fraction change indicating that the Type Ib/c fractions over redshift are good tracers for the metallicity evolution.

\begin{table}
    \centering
    \begin{tabular}{c|c|c|c}
    \hline
    & \textbf{II (IIP \& non-IIP)} & \textbf{Ib} & \textbf{Ic} \\ 
    & \multicolumn{3}{c}{CCSN Fraction} \\
    \hline
    \textbf{Empirical} & 0.71 & 0.17 & 0.11 \\
    \textbf{Millennium} & 0.73 & 0.15 & 0.12 \\
    \textbf{EAGLE} & 0.64 & 0.22 & 0.14 \\
    \textbf{TNG} & 0.67 & 0.20 & 0.13 \\
    \hline
    \textbf{\citet{shivvers_2017}} & 0.80 & 0.11 & 0.09  \\ 
    \hline
    \end{tabular}
    \caption{Fraction of CCSN subtypes at $z=0$ for predicted rates. \citet{shivvers_2017} is volume-limited to 60Mpc.}
    \label{tab:ccsn_subtype_fractions}
\end{table}


\subsection{Long Gamma-Ray Bursts}

\begin{figure}
    \centering
    \includegraphics[width=\linewidth]{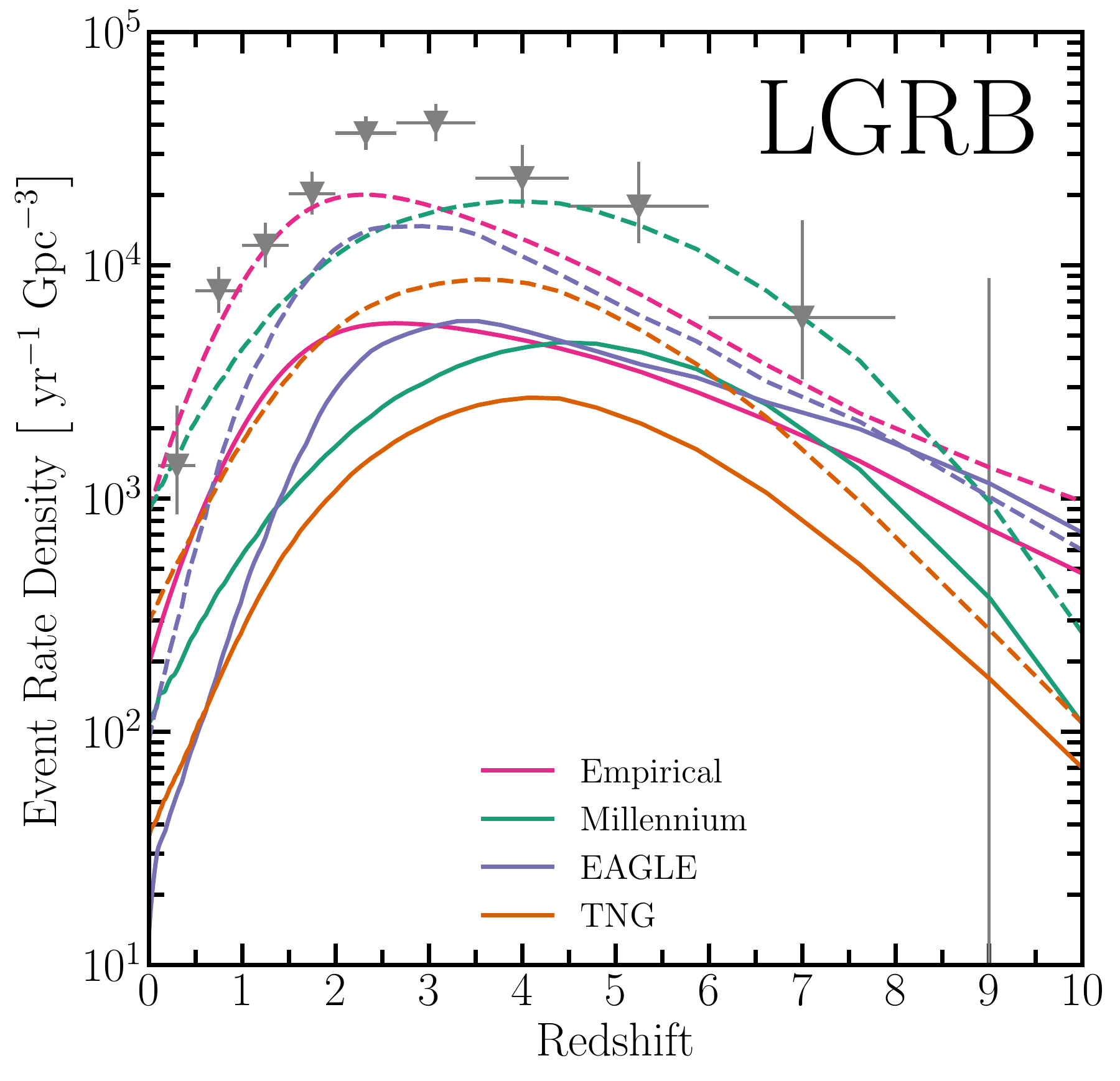}
    \caption{The predicted LGRB rates from the standard BPASS output (solid) with additional post-processed LGRB pathways added from \citet{chrimes_2020} (dashed) are shown for all four environment prescriptions.  The opening angle and luminosity corrected SHOALS rates are shown for comparison (grey triangles) \citep{perley_2016}.}
    \label{fig:LGRB}
\end{figure}

LGRB afterglows have been observed as coincident with broad-lined Type Ic (Ic-BL) and represent a subset of energetic Ic events, where a relativistic jet is launched from the surface of a nascent black hole or magnetar formed during core-collapse. The furthest GRB has been measured at $z=9.4$ \citep{cucchiara_2011}, but it is difficult to derive a volumetric rate for such sources because their emission is highly relativistic beamed. As a result, their observability depends on the event's geometry, specifically the jet opening angle. 

We use the observed GRB rate over redshift from the SHOALS sample \citep{perley_2016} that we correct for the event geometry and missed low-luminosity events. To achieve this, we adopt the method from \citet{chrimes_2020}, and integrate over the GRB luminosity function of \citet{pescalli_2016} from a isotropic equivalent energy of $\text{E}_\text{low} = 10^{48.1}$ erg to $\text{E}_\text{max} = 10^{56}$ erg, while correcting for an opening angle of $\theta = 9.9 \degree$. The grey triangle in Figure \ref{fig:LGRB} shows the corrected SHOALS rate. 

The predicted LGRB rates from BPASS, shown as solid lines in Figure \ref{fig:LGRB}, only contain events formed through chemically homogeneous evolution at low metallicities \citep{eldridge_2017}, which is lower than observed, as expected. Additional tidal formation pathways have been added post-process by \citet[][dashed line in Figure \ref{fig:LGRB}]{chrimes_2020}. These pathways mostly contribute at low redshifts and boost the predicted rate to a similar order of magnitude as the observations. However, the shape of the predicted rates no longer follows the same gradient as the observations and peaks at lower redshift. Due to the difficulty in constraining the observed rate and because the agreement is, of course, in part a consequence of the tuning of the LGRB opening angle parameters with a similar empirical CSFRD prescription as adopted here \citep{chrimes_2020}, it is not possible to distinguish between the star formation histories as of yet.

Although the LGRB are related to the Type Ic, their relative fractions evolve in opposite fashion in Figure \ref{fig:CCSN_fraction}. While the Type Ic increase over redshift, the LGRB rate drops significantly, because angular momentum required for the LGRB is removed by the stronger stellar winds in a more enriched Universe \citep{woosley_2002, vink_2001} and are therefore sensitive to low metallicity star formation. This relation between the two event rates can help us probe the metallicity distribution of a stellar population, especially since the chemically homogeneous LGRB event have a short delay allowing us to probe the change in low metallicity star formation.


\subsection{Pair Instability Supernovae}

\begin{figure}
    \centering
    \includegraphics[width=\linewidth]{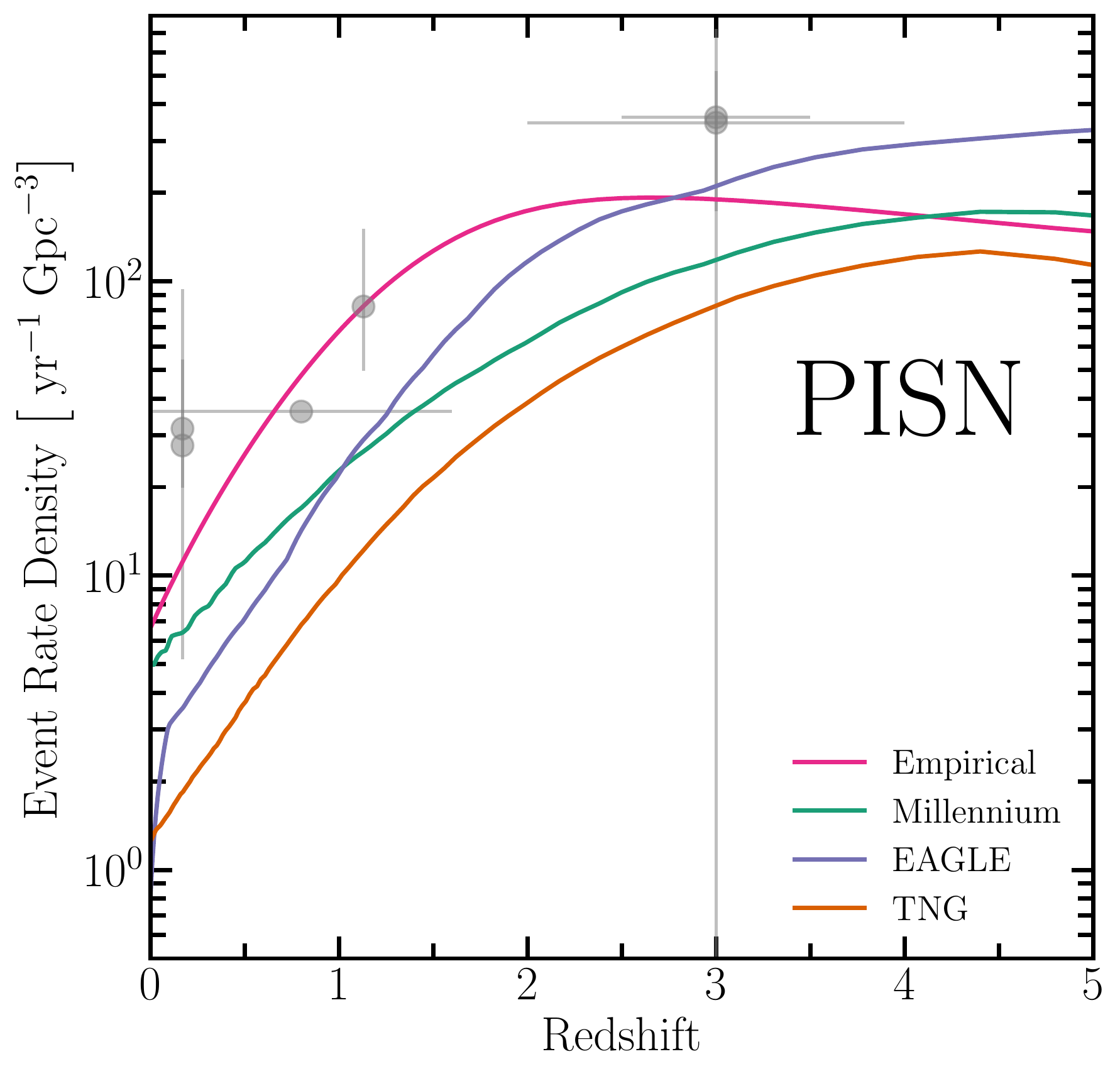}
    \caption{Predicted Pair-Instability Supernova rates with observational rates from Super-Luminous Supernova Type-I measurements (see appendix Table \ref{tab:SLSN_obs}.}
    \label{fig:PISN}
\end{figure}

Very massive stars ($\gtrsim 100$ M$_\odot$) reach the end of their life in only a few million years. The fast fusion leads to a low internal core density, while the temperature in the helium core reaches electron-positron production levels, causing the radiative pressure to drop due to the removal of photons. Without a force to counteract gravity, the star collapses in a PISN and completely disrupts the star, leaving no remnant behind \citep{fowler_1964, rakavy_1967, heger_2002}. The short lifetime, due to the high hydrogen burning rate, and the high metallicity sensitivity makes this event a perfect candidate for SFH probes \citep{yusof_2010, yusof_2013, dessart_2012, fryer_2001, eldridge_2019}.

Currently, no smoking-gun evidence of a PISN has been observed, although a few hydrogen-poor super-luminous supernovae (SLSN-I) have been identified as possible candidates \citep{woosley_2007, cooke_2012, gal-yam_2009, terreran_2017, gomez_2019}. Their energy requirements are too high to be consistent with the classical core collapse mechanism, but this remains unproven \citep{kozyreva_2015} and alternative explanations, such as magnetars \citep{howell_2017, kasen_2010, woosley_2010, inserra_2013}, rotational PISN \citep{renzo_2020a}, and late leakage from pulsar wind nebula \citep{dessart_2012}, are also likely. Nonetheless, in Figure \ref{fig:PISN} we have used the observed rate of SLSN-I that are possible PISN for comparison. The event rates from the simulations show the non-smooth nature of the low metallicity star formation rate at low redshift, specifically the EAGLE simulation. Its event rate has a significant drop at $z=0.1$ which is caused by a drop in the low metallicity star formation.
Since they have limited star formation at low redshift and PISN mostly occur in low metallicity environments, only the predicted event rate from the empirical CSFRD aligns with the observations. The uncertainty in the formation pathway of SLSN-I makes it not possible to state if SLSN-I are a good indicator for the PISN rate, except that their observed rate is within an order of magnitude of our predictions.

\subsection{Compact Objects} \label{subsec:CO}

\subsubsection{Binary Black-hole mergers} \label{subsubsec:BBH}

\begin{figure}
    \centering
    \includegraphics[width=\linewidth]{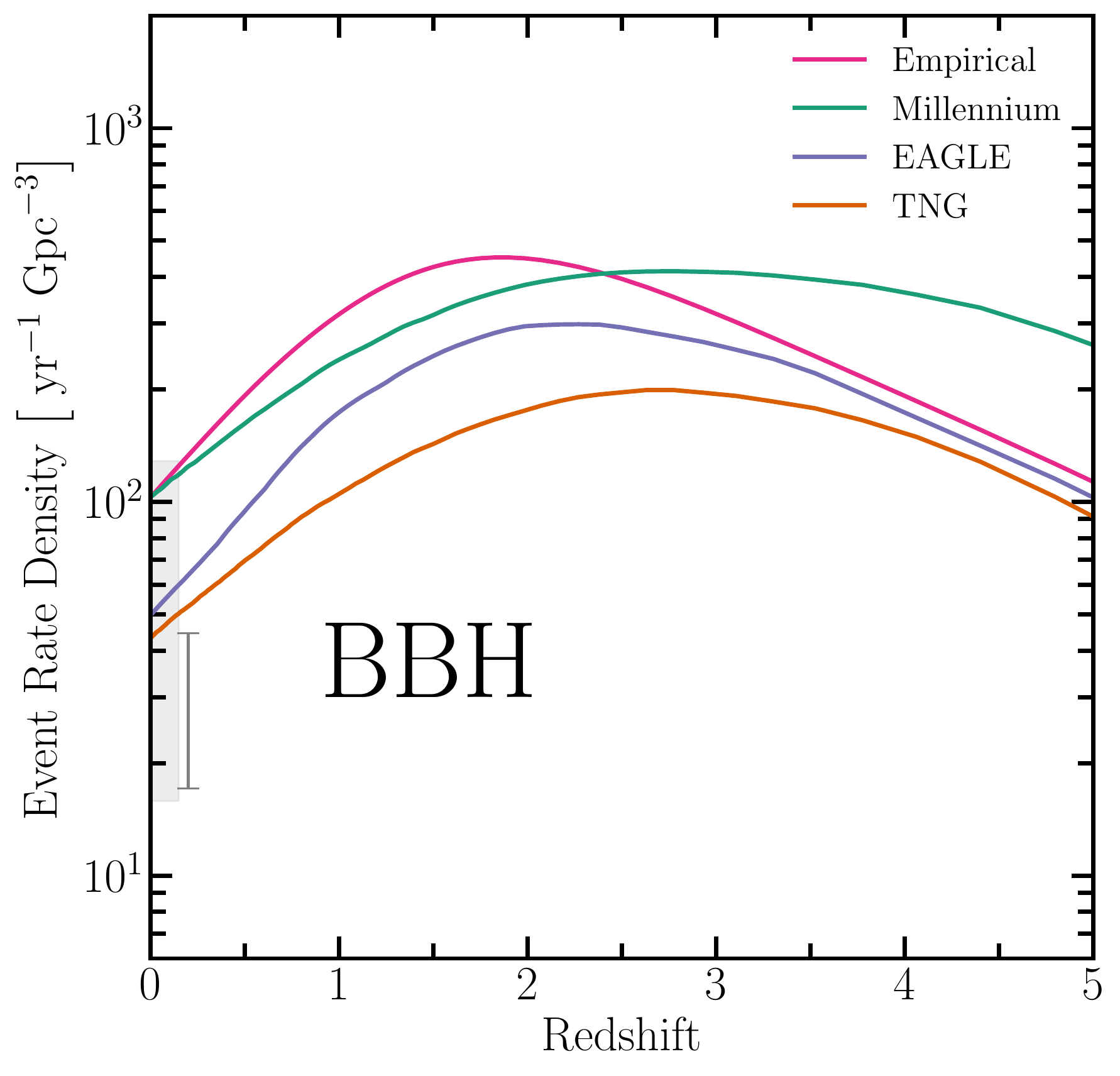}
    \caption{Binary BH merger rate predictions with observations from \citet{abbott_2021c} with the lowest 5\% to highest 95\% credible interval of their considered population models at $z=0$ and $z=0.2$ for the non-evolving and evolving merger rates, respectively. The $z=0$ rates are shaded and extended out to $z=0.15$ for clarity.}
    \label{fig:BHBH}
\end{figure}

The binary nature of compact object mergers leads to the domination of events with long delay times, which would result in the peak of BBH mergers occurring after the peak star formation history. However, since the BBH merger rate is sensitive to metallicity and mostly occurs at low metallicity, this does not have to be the case. As we see in Table \ref{tab:peaks} and in Figure \ref{fig:BHBH}, only the TNG and Millennium predictions have their peak after the maximum SFR, while the empirical and EAGLE predicted BBH rates peak before. This indicates that most of the BBH events originate from low metallicity populations in the earlier Universe as expected.

\begin{figure*}
    \centering
    \includegraphics[width=\linewidth]{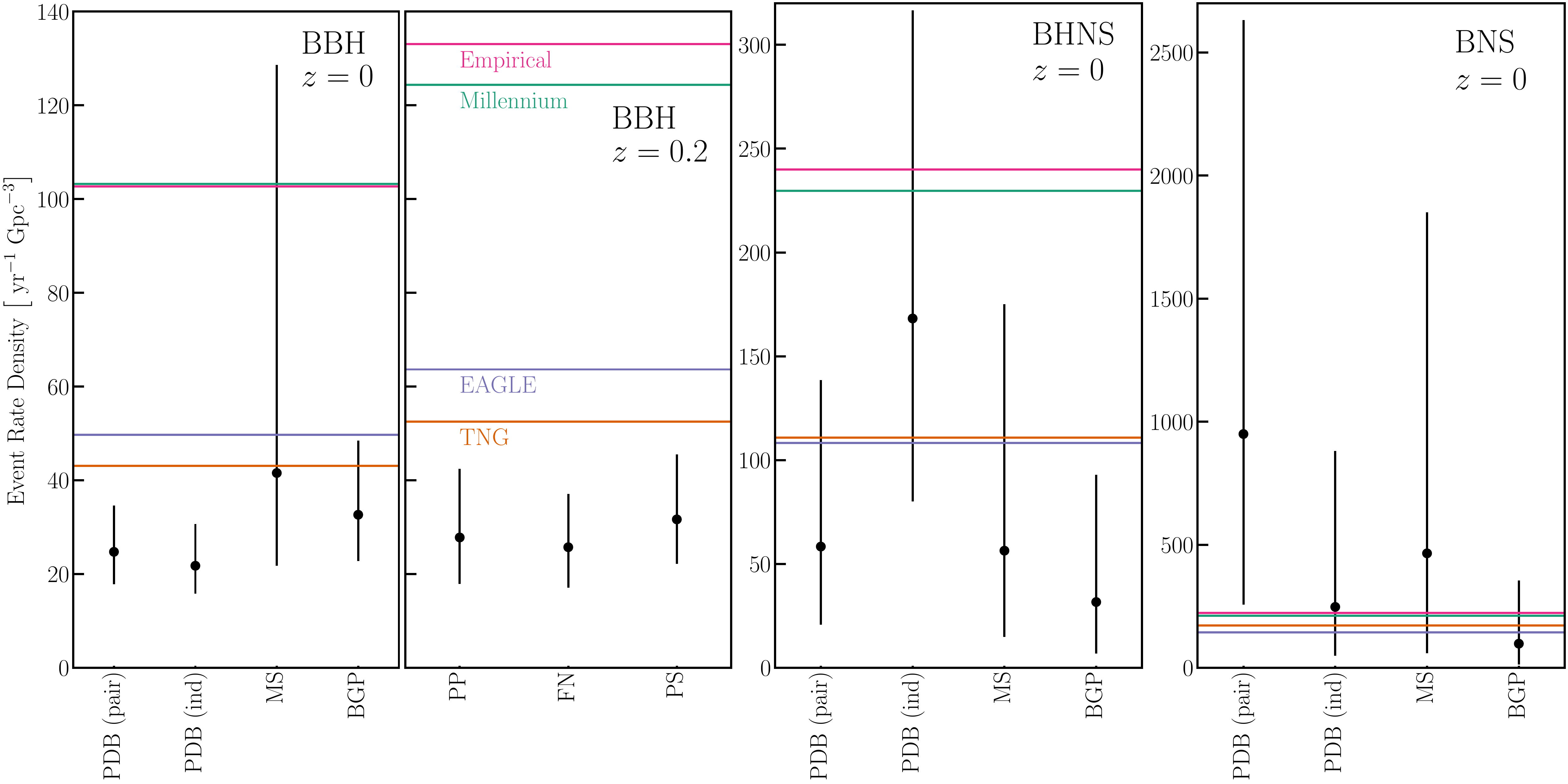}
    \caption{Observed BBH, BHNS, BNS merger rates from different assumed population models by \citet{abbott_2021c}. The BBH rates are given shown at $z=0$ for the non-changing distributions, while the credible range for the changing distributions is shown at $z=0.2$, where their error is the smallest.}
    \label{fig:GW_rates}
\end{figure*}

A more direct comparison is possible with recent results from GWTC-3, which includes estimates for a variety of population models at $z=0$ for non-evolving merger rates and at $z=0.2$ for redshift dependent rates \citep{abbott_2021c}. Figure \ref{fig:BHBH} shows their range when only considering the lowest 5\% and highest 95\% credible boundaries out of the \texttt{PDB (ind)}, \texttt{MS}, and \texttt{BGP} models \citep[For a description of the models see][]{abbott_2021c}, whose ranges are shown in Figure \ref{fig:GW_rates}. All predictions fall within the combined 90\% credible interval from $16$ to $130$ yr$^{-1}$ Gpc$^{-3}$ for the non-evolving merger rate at $z=0$, although the Millennium and Empirical predictions are at the higher end of the observational range. However, the merger rate increases over redshift and when this is taken into consideration the observed rate decreases to $17.3$-$45$ yr$^{-1}$ Gpc$^{-3}$, as the bar at $z=0.2$ in Figure \ref{fig:BHBH} shows. This combined credible interval has been constructed in similar fashion to that at $z=0$, but considers three BBH population models,\texttt{PP}, \texttt{FN}, and \texttt{PS} \citep[For a description of the models see][]{abbott_2021c}, which evolve over redshift. At $z=0.2$ the observational constraints from \citet{abbott_2021c} are the strongest and the combined 90\% credible range from their collection of models lies below the predicted merger rates, despite that the simulations with a fast enrichment, the EAGLE and TNG simulations, approach upon this range, which shows the strong correlation between the BBH rate and metallicity evolution.

More detail can be obtained by looking at the specific population models used by \citet{abbott_2021c} instead of looking at the combined credible interval. The conversion from measured to intrinsic rates introduces several model-dependent uncertainties in the BBH rate and can move the observed rate both towards and away from our predictions. For the constant merger rate at $z=0$, shown in the left-most panel in Figure \ref{fig:GW_rates}, the majority of uncertainty in the credible interval comes from the \texttt{Mixed-Source model}, while the other models are clustered around the same rate of $\sim30$. The predicted rates from the empirical prescription and Millennium simulation are an order of magnitude higher than these observed rates. Only the TNG and EAGLE simulation are close to this range, but still overestimate the predicted BBH merger rate compared to the \texttt{PDB (ind) model} by a factor 2.0 and 2.2, respectively. The fact that the TNG simulation also overestimates the fiducial evolving observed rate (\texttt{PP}) by 1.9, indicates a systematic overprediction. This might be an effect of the details of the assumed populations compared to BPASS, where more low mass black holes are formed \citep{ghodla_2021}, thus resulting in our predicted rates for BBH mergers being too high.

On the other hand, \citet{abbott_2020c} also found that the BBH rate increases with a factor $2.7_{-1.9}^{+1.8}$ between $z=0$ and $z=1$, which lies between the predicted values from the  Millennium ($2.3$), TNG ($2.4$), empirical ($3.1$), and EAGLE ($3.5$) predictions, indicating that the rate only requires a small downward adjustment to within the observed range that can be achieved by the following methods:

First, the BBH merger rate is sensitive to the environmental parameters, such as the star formation rate and metallicity evolution in the early Universe \citep{dominik_2013, mapelli_2017, lamberts_2018, mapelli_2019, santoliquido_2020a, artale_2020, tang_2020, neijssel_2019}, where the differences between SFH in our simulations are most apparent. This effect is already apparent in the predicted rates with the TNG and EAGLE simulation having a faster enrichment and lower BBH rates. An even faster increase in mean metallicity could reduce the BBH cosmic rate further, but this would also increase the overpredicted Type Ib/c fraction and further underestimate the LGRB rate. Although the constraints on the observed fraction and LGRB rate are limited, we should also focus on other influences on the cosmic BBH merger rate, such as the assumed constant binary fraction over redshift and metallicity. For close binary systems with solar-type stars, the binary fraction decreases significantly with metallicity \citep{moe_2019}. A similar relation might hold for massive stars, but the binary fraction is difficult to infer from early Universe observations.

The second area of interest is the physics and parameters assumed in the stellar evolution models. While BNS merger rate is most sensitive to these parameters, specific processes could contribute to a lower BBH rate but unchanged BNS rate. For example, altering the prescription used to predict SN outcomes \citep{dabrowny_2021} or including Pulsation Pair-Instability \citep{belczynski_2016a, farmer_2019, stevenson_2019} can decrease the compact remnant masses, making it easier for systems to become unbound, thus, possibility reducing the BBH and BHNS rate. In the case of the latter, the pulsation pair-instability can even prevent black holes from forming. Other possibilities include a higher natal kick, a different common envelope prescription \citep{santoliquido_2020a, dubuisson_2020, marchant_2021}, or stronger stellar winds at low metallicity \citep[][and references therein]{mapelli_2021a} can be implemented. However, this would alter the BNS rate, while not significantly influencing the BBH rate.

Finally, including GW190814 in the BBH merger rate can drastically change the observed intrinsic rate. For example, in \citet{abbott_2020c} the inclusion of GW190814 changes the observed rate in our cosmology to $54_{-29}^{-53}$ Gpc$^{-3}$ yr$^{-1}$, due to an increase in the expected number of low mass black holes resulting in nearly the same rate as predicted by the EAGLE and TNG simulations at $z=0$. The nature of GW190814, however, is an area of active discussion and the models by \citet{abbott_2020c} do not extrapolate well to the GW190814 masses ($\text{M} < 3\msol$).

\subsubsection{Binary Neutron-stars mergers}

\begin{figure}
    \centering
    \includegraphics[width=\linewidth]{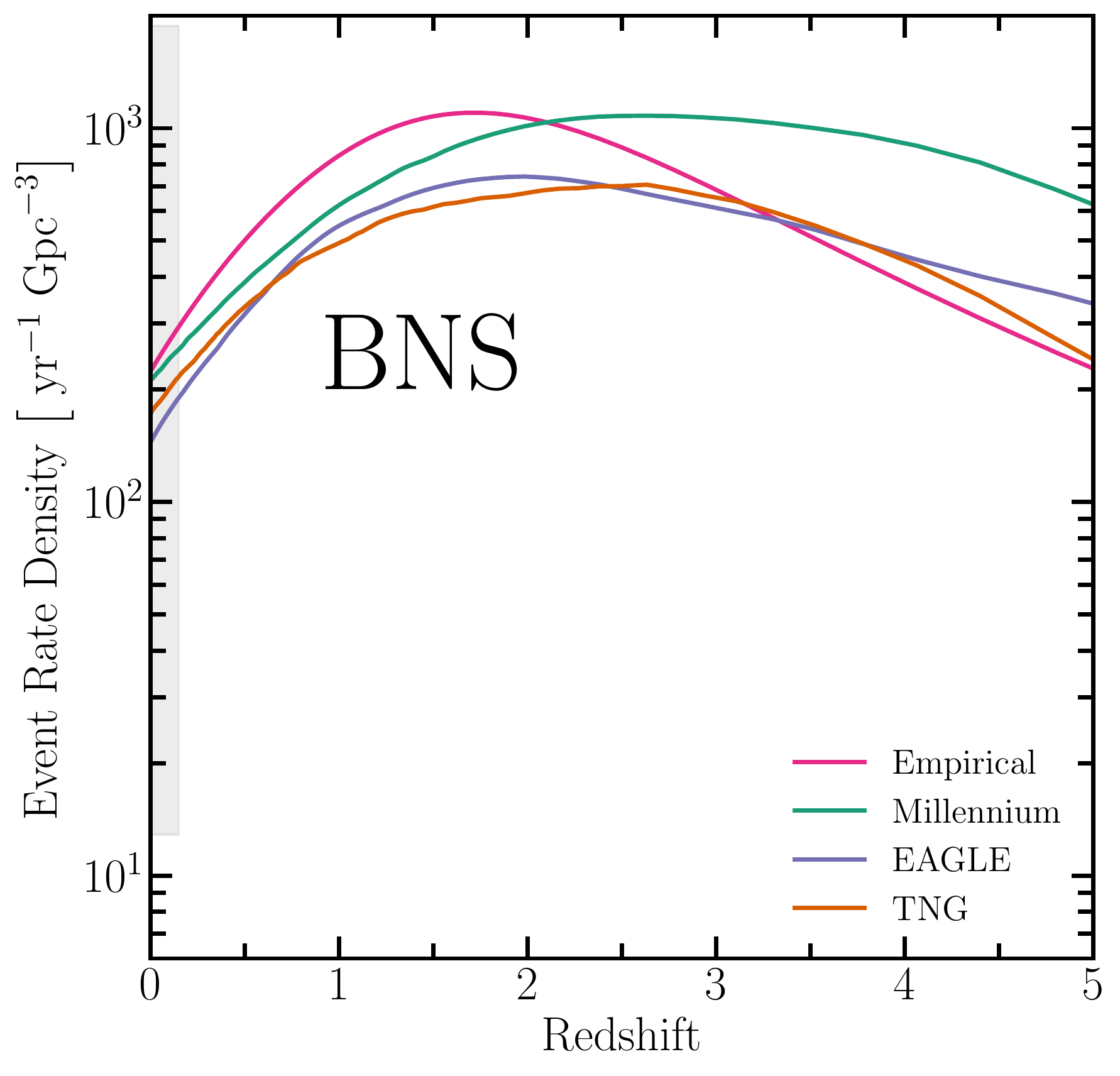}
    \caption{Binary Neutron Star merger rate with maximum credible range at $z=0$ from \citet{abbott_2021c}.}
    \label{fig:NSNS}
\end{figure}

Like the BBH population, the BNS rate has been directly measured at $z=0$ using gravitational wave measurements. While short gamma-ray burst can provide observations at higher redshift, the conversion from measurement to intrinsic rate suffers from a similar dependence on opening angle as LGRB. The intrinsic merger rate is mostly independent of metallicity \citep{tang_2020, giacobbo_2018}, but is highly sensitive to the natal kick and Common Envelope Evolution prescription \citep{santoliquido_2020a, dominik_2013, giacobbo_2018}. Figure \ref{fig:NSNS} shows that all four predictions lie within the maximum credible region at $z=0$, which is a positive sign for the implemented physics model. Breaking down the measured interval further in Figure \ref{fig:GW_rates}, we see that the rates fall within the boundaries of most populations models considered by \citet{abbott_2021c}, expect for the \texttt{PDB (pair) model}, which has a higher BNS merger rate than any of the models predict. Our predictions agree with the other population models, but no environmental prescription is preferred by the BNS rate, which is as expected since it is mostly independent of environmental parameters. Instead better constraints can be placed by looking at the BNS chirp mass distribution and surviving BNS systems, however such comparisons go beyond the scope of this paper. 

\subsubsection{BH-NS mergers}

\begin{figure}
    \centering
    \includegraphics[width=\linewidth]{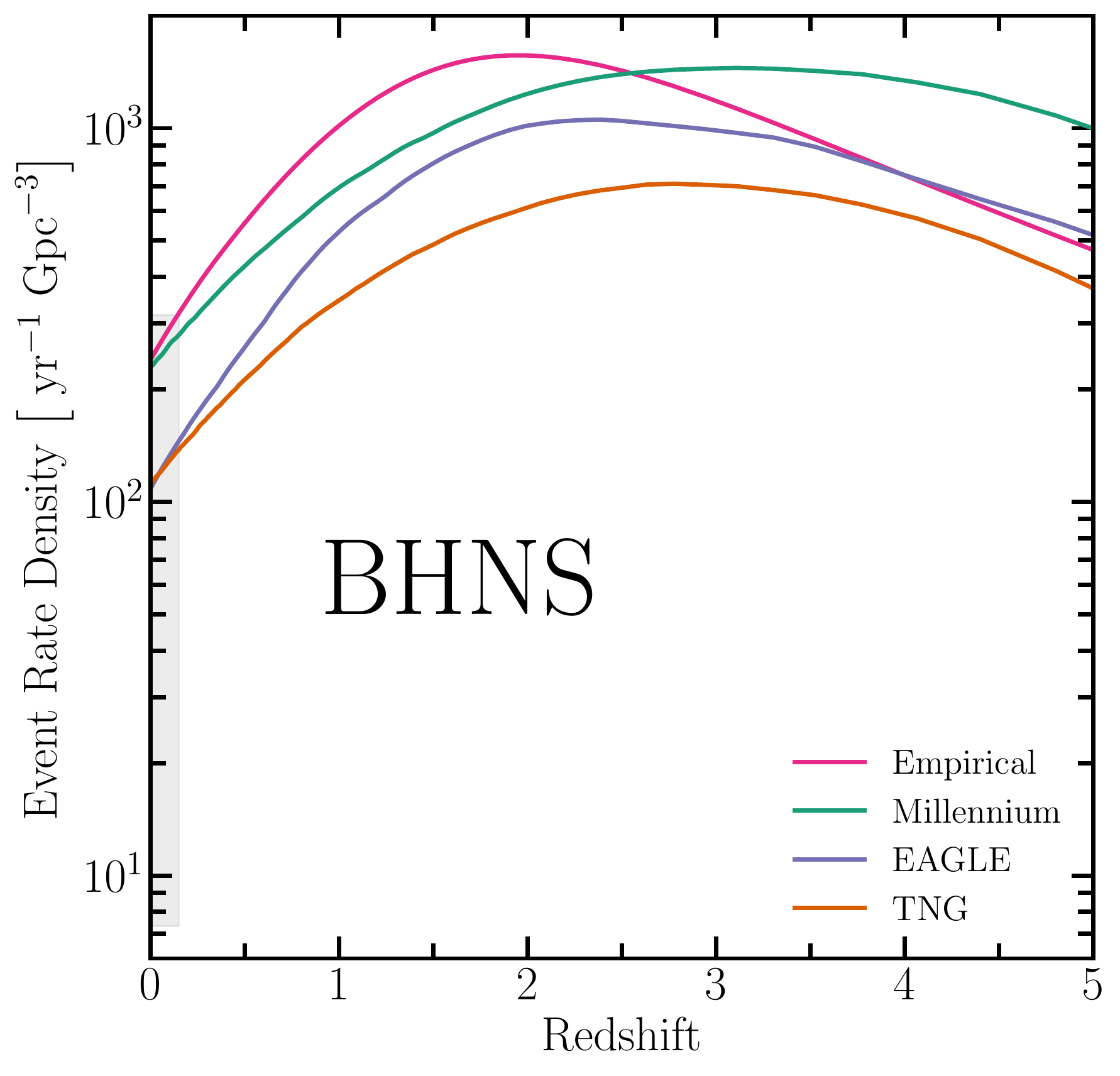}
    \caption{The observed maximum credible range of the BHNS merger rate at $z=0.2$ from \citep{abbott_2021c} compared against the TNG, EAGLE, Millennium, and empirical predictions.}
    \label{fig:BHNS}
\end{figure}

After observations from \citet{abbott_2021} produced a weak constraint on the BHNS merger rate, \citet{abbott_2021c} improved upon this by using a joint-analysis of the BBH, BHNS, and BNS populations with multiple population models. As shown in Figure \ref{fig:BHNS}, all predictions lie within the maximum credible region of the models considered, although similar to the BBH merger rate, the empirical and Millennium predictions are at the higher end of this range. The majority of the uncertainty in the rate, this times comes from the \texttt{PDB (pair) model}, as can be seen the second panel to the right of Figure \ref{fig:GW_rates}. Similar to the BBH rates, the EAGLE and TNG prediction, are better estimators of the observed rate, which is most likely a result of the low metal-poor star formation rate, since black hole formation is dependent on the metallicity of the star formation environment \citep{mapelli_2021a}. However, the difference is more pronounced in the BBH rate than the BHNS rate, which agrees with the finding of \citet{broekgaarden_2021} that show a stronger influence of the stellar physics on the BHNS rate than the star formation environment. The interplay between the cosmic BBH, BHNS, and BNS rates is essential in finding the exact origin of the high black hole formation estimation, but requires further investigation into the influence of the remnant mass determination, mass transfer efficiency, and natal kick.

\section{Discussion} \label{sec:discussion}

\subsection{Combined rates analysis} \label{subsec:combined_rates}

Individually, each transient rate prediction only provides limited information about the different factors that affect it, but taken together it becomes possible to disentangle these influences. To this end, we have performed a reduced $\chi^2$ calculation on the electromagnetic event rates with observation available expect the LGRB rates. The reason for this exclusion is the fact that the geometrical parameters for the observed rate normalisation are fitted using the same metallicity distribution and a similar CSFRD as our empirical prescription. They will therefore be biased towards the empirical CSFRD. 

Looking at the combined reduced $\chi^2$ of the electromagnetic transients in Table \ref{tab:chi2}, the TNG simulation has the closest $\chi^2$ to 1 with a value of 0.86, followed by a reduced $\chi^2$ of 0.74 by the EAGLE simulation, while the empirical and Millennium predictions have reduced $\chi^2$ of 0.58 and 0.51, respectively. Although the TNG simulation provides the best match to the data, the limited difference in reduced $\chi^2$ values do not allow us to distinguish between the predictions.

The compact objects are excluded in the $\chi^2$ calculation, because there is no robust method to included the credible intervals without knowing their distributions. Instead, we look at these rates qualitatively by considering the population models in \citet{abbott_2021c} with each having its advantages and disadvantages that can drastically alter the observed rates. For example, the fiducial BBH population model (\texttt{PP}) evolves over redshift, but does not consider the NS masses as part of the same population. On the other hand, the \texttt{PDB (ind)} model fits the BH and NS masses, but assumes a non-evolving merger rate over redshift. To show the influence of these assumptions, we showed in Figure \ref{fig:GW_rates} the rates for each population model and compared them against the predicted GW rates in Section \ref{subsec:CO}. Here, we consider all gravitational wave transients simultaneously and, while the BNS does not provide additional constraints due to the limited observations and, thus. the wide credible interval, we see that the Millennium simulation and empirical prescription overestimate the BBH and BHNS merger rates for the majority of the population models. The EAGLE and TNG predictions, on the other hand, best approximate the observed BBH, BHNS, and BNS merger rates independently of the used population model. This difference is most likely caused by the metallicity distributions and differences in CSFRD. The fact that even these BBH predictions are outside the 90\% credible interval for some population models indicates that the BBH rate is overestimated by BPASS, as discussed in Section \ref{subsubsec:BBH}.

Because the CSFRD of the Millennium simulation aligns with SFR observations at $z<2$, short delay time events, such as PISNe and CCSNe, are reasonably well predicted at these redshift. However, at higher redshift the CSFRD does not align with SFR observations and these estimated rates deviate from the other predictions. This is not taken into account into our $\chi^2$ calculation because of limited rate observations in this regime. For long delay time events, the influence of the CSFRD combines with the metallicity distribution. Metallicity independent event types, such as Type Ia and BNS, align with observations, indicating that the total amount of stellar material formed in the past is correct, but metallicity dependent events, like the BBH and BHNS, show that the SFR at early times, when there most low-metallicity star formation takes place, is too high resulting in overestimated cosmic rates. This is further motivated by the short delay time and metallicity dependent LGRB rate, whose shape over redshift does not align with that of the observations and peaks at an early redshift.

The empirical CSFRD solves the SFR observation misalignment, but still has a significantly different metallicity distribution over redshift than the simulations, as shown in Figure \ref{fig:Z_dist}. Its slow metallicity increase over redshift leads to the overestimation of the BBH and BHNS merger rates, since the low-metallicity star formation at early times is a primary indicator for the BBH rates \citep{neijssel_2019}. The EAGLE and TNG simulations, on the other hand, are able to better estimate the observed BBH and BHNS rates due to faster enrichment in the early Universe, even when considering an evolving BBH merger rate. Although subject to model uncertainties, this means that more detailed observations of the BHNS and BBH rates can be used to constraint the metallicity-specific CSFRD at high redshift, where observations of metallicity-specific SFRs are limited, providing us with a more complete understanding of the Universe.

\subsection{Caveats in the estimations} \label{subsec:discussionrates}

While the EAGLE and TNG simulations are able to reduce the BBH and BHNS rates compared to the empirical and Millennium CSFRD predictions, the estimates remain high, since only isolated binary formation is considered. In reality, a mixture of isolated systems and dynamical interactions in globular clusters, young stellar clusters, nuclear clusters, isolated triples, and systems in active galactic nuclei disks  will contribute to the total cosmic merger rates \citep[][and references therein for dynamical interactions]{zevin_2021, santoliquido_2020a, gallegos-garcia_2021, bouffanais_2021}.
Consequently, even when considering the EAGLE and TNG simulations, the BBH and BHNS rates are high and have to be reduced further. This can be achieved by altering the natal kick, mass transfer efficiency, or common envelope prescription, as described in Section \ref{subsubsec:BBH}, and possible differences in mass distributions could identify areas of improvement \citep{ghodla_2021,mapelli_2019}. However, we would like to point out that in BPASS v2.2.1, the merger time for compact objects is calculated using the mean of a collection of systems instead of each individual system. This could result in more systems merging within the Hubble time that otherwise would not. The individual calculation will be implemented in a future version of BPASS.

Besides the overestimation of the BHNS and BBH rates, the EAGLE and TNG simulations have high Type Ib/c fractions at $z=0$ compared to the empirical and Millennium predictions due to the higher mean metallicity resulting in stronger stellar winds \citep[e.g.][]{vink_2001}. Moreover, they also are significantly higher than observed by \citet{shivvers_2017} and \citet{perley_2020} indicating too high a typical metallicity at low redshift. Additionally, the short delay-time low-metallicity dependent PISN and LGRB rates are underestimated compared to observations, and from the lower Type Ib/c fraction from the Millennium and empirical CSFRD, which have a lower mean metallicity near $z=0$. Together, these results indicate that either the metallicity is too high at these redshifts, or the strength of the stellar winds is too high, both resulting in more Type Ib/c and less PISN and LGRBs. It is, however, important to mention that available observations for the Type Ib/c fractions, PISN and LGRB rates are currently limited, but that the continuation of surveys from the Zwicky Transient Facility \citep{perley_2020} and ATLAS \citep{smith_2020b} will provide better constraints on the Type Ib/c fractions and SLSN observations=, while next generation observational facilities, such as THESEUS, will be able to detect large samples of GRBs at $z>6$ \citep{tanvir_2021}.

\begin{figure*}
    \centering
    \includegraphics[width=\linewidth]{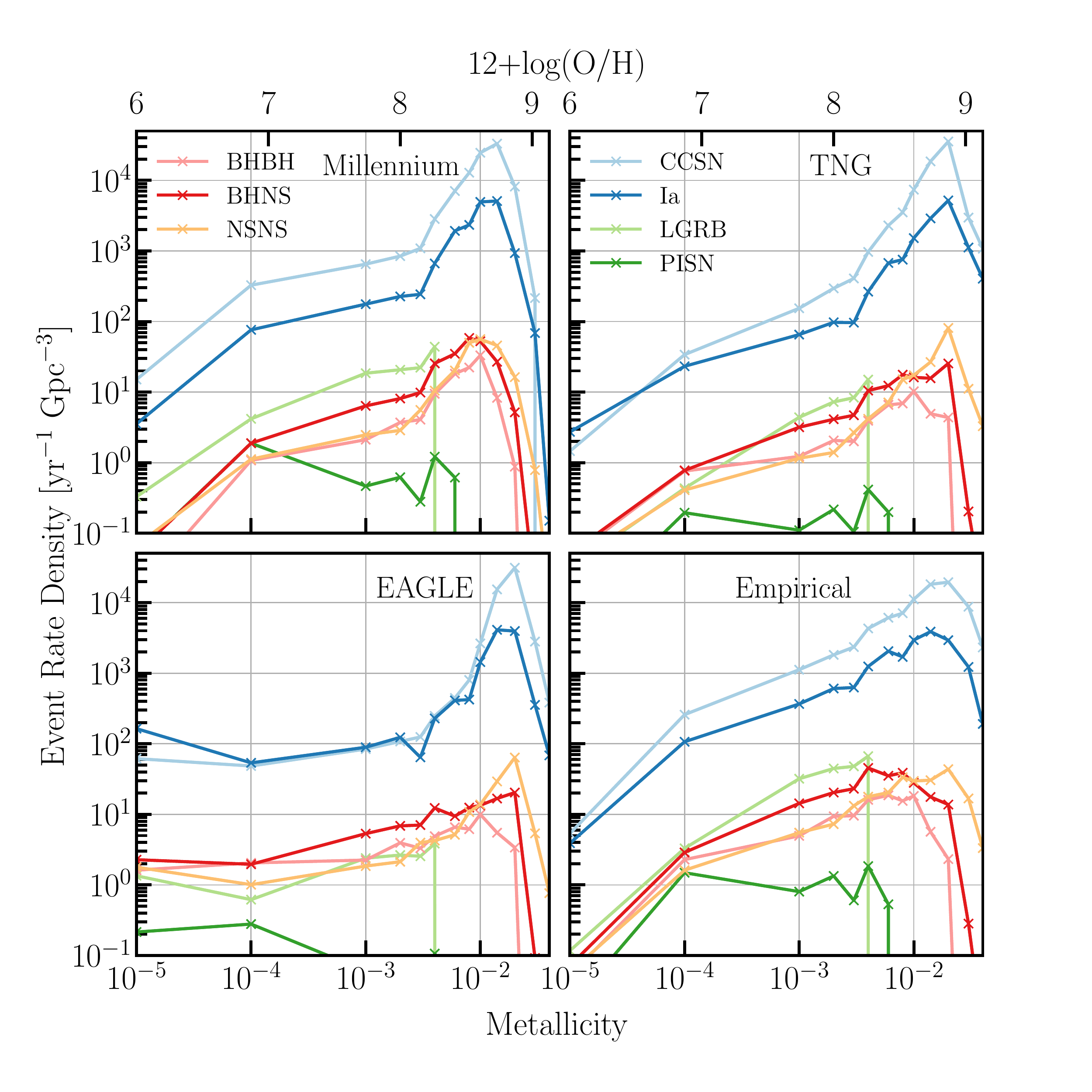}
    \caption{The event rate distribution at $z=0$ over metallicity for each of the four environment prescriptions. The transformation from metallicity to the O/H ratio is performed using the method in \citet{chrimes_2020}.}
    \label{fig:ER_over_Z}
\end{figure*}

With these future observations, it might become possible to distinguish between the similar EAGLE and TNG predictions and provide us with more understanding of the metallicity at high redshift. Moreover, the distributions of event rates at $z=0$ over metallicity, shown in Figure \ref{fig:ER_over_Z}, is another method to constrain the CSFRD and metallicity evolution. The distributions are all relatively similar, except for the EAGLE simulation, which stands out due to relative high event rates originating from regions of star formation with a metallicity below $Z=5 \times 10^{-3}$. A possible explanation might be a larger and more equal spread of star formation over metallicity in the EAGLE CSFRD, although Figure \ref{fig:Z_dist} does not provide a clear answer to this question, since the distributions of the EAGLE and TNG look similar at low redshift.

\subsection{Uncertainties}

There are three main sources of uncertainty in this work: the interference of volumetric rates from observations, the cosmological simulations, and the stellar modelling. We have already touched upon a variety of uncertainties in observations, since they are essential in testing our predictions, but observations at high redshift remain limited and complex, although future surveys will allow for better constraints.

The stellar models used in creating the DTDs used in their study are from BPASS. The uncertainties inherent in the assumed physics within these stellar models and the population synthesis undertaken to combine them is common to all such codes. For example, within BPASS, the Type Ia rate is dominated by single degenerate events, while \citet{ruiter_2009} and \citet{mennekens_2010} predict the double degenerate channel to be dominant. The balance between these pathways depends on mass-loss rates, metallicity, and binary fraction. \citet{ruiter_2009} only looks at a metallicity of $Z=0.02$ and a binary fraction of 50\%. \citet{mennekens_2010} uses the same metallicity, but with a binary fraction of 100\% and alterations to the mass transfer efficiencies. Moreover, BPASS does not include the effects of magnetic wind braking, and its inclusion might shift the Type Ia rate \citep{eldridge_2017, stanway_2018a}. These inherent uncertainties in the assumed physics of the stellar models propagate to uncertainties in the predicted transient rates. Nonetheless, we note that the BPASS results have been validated and tested against a wide range of observations beyond just astrophysical transients providing confidence in the predicted DTDs and they are therefore sufficiently accurate to study the large scale trends presented in this work \citep[][and references therein]{eldridge_2017, stanway_2018a} .

The final area of uncertainty comes from the cosmological simulations. Due the age of the semi-analytical Millennium simulation, it is unable to reproduce several observational constraints \citep{croton_2006, oliver_2010, wang_2019,lu_2014} and the cosmic rates. The newer hydrodynamic simulations, on the other hand, are able to reproduce a number of observed distributions \citep[See][]{schaye_2015, crain_2015, springel_2018, nelson_2018, pillepich_2018a, naiman_2018, marinacci_2018} and have better estimates for the cosmic transients rates due to the presence of a SFH and metallicity evolution per galaxy. These are significant benefits of the cosmological simulations compared to the empirical prescription, but they do have some drawbacks. The detail of the large scale cosmological simulations does not provide information about the structure and metallicity distribution within the galaxy limiting us to the average metallicity of the galaxy. In reality, the centre and outer regions of galaxies have different abundances \citep{metha_2020, metha_2021}, primarily impacting the metallicity dependent rates.

Furthermore, the metallicity evolution of the cosmological simulations is formed through enrichment processes, such as supernova feedback. To account for these processes, internal definitions for different supernova types are assumed. For the TNG simulation, these align with observations as shown by \citet{naiman_2018}, but are different than predicted by the BPASS models. While the Millennium and EAGLE simulations do not have explicit SN rates, they do use prescriptions that depend on the fraction of massive stars ending their life in a stellar explosion by which they determine the metallicity evolution and stellar growth of galaxies in the simulations \citep{katsianis_2017, croton_2006}. The predictions using the BPASS models can lead to a different stellar rate, as with the TNG simulation, and therefore alter the enrichment in the simulation. Thus, for a self-consistent model, the SNe feedback would have to be modelled using the BPASSS supernova rates.

There are two more elements of the simulations that influence the outcome of the predicted rates. First of all, the internal star formation in the TNG, EAGLE, and Millennium simulations assume a Chabrier IMF \citep{delucia_2007,mcalpine_2016, pillepich_2018a}, while the BPASS DTDs used originate from a Kroupa IMF. This discrepancy could result in a deviation in the predicted rates, but the difference would be small since the observational correction from luminosity to SFR between the Kroupa and Chabrier IMF is minimal \citep{madau_2014}.
The second internal component is the assumed cosmology in the simulation. While some hydrodynamic simulations are completely scale free and independent of the Hubble parameter, more sophisticated simulations require absolute values for processes, such as cooling and SN feedback \citep{croton_2013}. These dependencies cannot be removed as we have done with the volume and will introduce variations in the predicted rates due to the cosmology. However, its influence is small, especially since the cosmologies from the EAGLE and TNG simulations are similar to the assumed value in this paper. Since the Millennium simulation is a semi-analytical model, it does not suffer the same dependence.

\section{Conclusion} \label{sec:conclusion}

In this paper, we have predicted electromagnetic and gravitational wave cosmic transient rates using detailed stellar models from BPASS and four prescriptions of the star forming environment from well-known cosmological simulations. These include an empirical prescription and three numerical models originating from the Millennium, EAGLE, and Illustris-TNG simulations, which provide detailed SFH and metallicity evolution for each simulated galaxy over the history of the Universe.
These additional details lead to significantly different cosmic transient rate prediction, which we compared against observations and against each other, focusing on the difference between the simulations and the empirical prescription.

\begin{enumerate}
    \item There are significant differences between the predicted rates from the empirical model and those from the cosmological simulations. The predicted rates are most different when the delay-times distribution for the events are extended and/or when the rates are highly metallicity dependent. This result suggests that care should be taken in choosing which cosmic star formation history to use, especially how the metallicity evolution is modelled when predicting transient rates. Those that are most sensitive are events with long, $>1$~Gyr, delay times and strong dependence on metallicity.

    \item The cosmological simulations have metallicity-specific CSFRD with reduced early star-formation and faster enrichment than the empirical prescription. Of the SFH considered, the Millennium simulation has the most uniquely shaped CSFRD, which does not agree with SFR observations, and has a near constant mean metallicity over redshift resulting in distinct cosmic transient rates. Compared to observations, the predictions from the Millennium simulation agree with the observations for the Type Ia, CCSN, PISN, and BNS rates, but overestimate the BHNS and BBH rates.
    The TNG and EAGLE simulation solve this overestimation with an increasing mean metallicity evolution and a observationally constraint CSFRD resulting in BHNS and BBH rates closer to observation irrespective of the assumed BH and NS population model, while at the same time the Type Ia, CCSN, and BNS rates are minimally affected. However, the PISN and LGRB rates do decrease significantly due to the higher metallicity, but the number of observations are limited or hard to constrain, although it might indicate that there is not enough low-metallicity star formation near $z=0$ or the used stellar wind prescription is too strong.

    \item We find that the predictions of the empirical prescription, based on the CSFRD from \citet{madau_2014} and metallicity evolution from \citet{langer_2006}, aligns well with observed CCSN and Type Ia rates from multiple surveys. Moreover, the predicted fraction of Type Ib/c at $z=0$ of 0.71 is similar to 0.722 found by \citet{perley_2020}, and the predicted LGRB, PISN, and BNS rates align well, although the observations for the LGRB and PISN are not well constrained. Furthermore, the BHNS and BBH rates are significantly overestimated compared to observations up to almost an order of magnitude. 
    
    \item The additional detail in the metallicity-specific CSFRD provided by the TNG and EAGLE simulations result in good cosmic rate estimations across the board, but especially improves upon the BHNS and BBH rates compared to the standard empirical prescription. These new cosmological simulations have been improved to fit a variety of observations. The semi-analytical models from the Millennium simulation, on the other hand, are older and unable to match the observed CSFRD over redshift. Together with the near flat metallicity evolution, it results in significant overestimation of the BBH and BHNS rate with minimal changes in the other rates compared to the empirical prescription.

    \item In this paper we have only considered isolated binary evolution for the BBH and BHNS rates. The Universe has more than one way to make compact object mergers happen with the contribution of each pathway, such as isolated and dynamic formation still being undetermined. This means that even the better rate estimations from the EAGLE and TNG simulations for the BBH and BHNS rates are likely still too high. Adjustment of the CEE or natal kick, or inclusion of pulsation pair instability  might be necessary for more accurate rate predictions.
    
\end{enumerate}

All together, we find that the EAGLE and TNG simulation provide the best metallicity-specific CSFRD based on the predicted cosmic rates for electromagnetic and gravitational wave transients. The additional detail provides clear benefits over the empirical prescription, closer matching the irregular and complex metallicity and SFR evolution of the real Universe, constraining environmental and evolutionary parameters. As the observational constraints improve over the coming decades for the star formation history and cosmic transients rates, we expect the true complexity of their variation over redshift to be revealed. 

\section{Acknowledgements}
MMB, JJE and HFS acknowledge support by the University of Auckland and funding from the Royal Society Te Apar\={a}ngi of New Zealand Marsden Grant Scheme. ERS acknowledges support by the University of Warwick and funding from the UK Science and Technology Facilities Council (STFC) through Consolidated Grant ST/T000406/1. We are grateful to the developers of python, matplotlib \citep{hunter_2007}, numpy \citep{harris_2020a} and pandas \citep{reback_2020}.

The BPASS project makes use of New Zealand eScience Infrastructure (NeSI) high performance computing facilities. New Zealand's national facilities are funded jointly by NeSI's collaborator institutions and through the Ministry of Business, Innovation \& Employment's Research Infrastructure programme.

\section{Data Availability}
Code and data products underlying this work are available at the organisational github \footnote{\url{https://github.com/UoA-Stars-And-Supernovae}} and website \footnote{\url{www.bpass.auckland.ac.nz}}.




\bibliographystyle{mnras}
\bibliography{references}




\appendix

\section{Observations}

\begin{table*}
\begin{center}
\begin{tabular}{c|c|c|c|c}     
\textbf{Redshift} & \textbf{Rate} & \multicolumn{2}{c}{\textbf{Uncertainty}}  & \textbf{ Reference} \\
& \multicolumn{3}{c}{[$10^5$ $h^{3}$ yr$^{-1}$ Gpc$^{-3}$]} & \\
\hline
0 & 0.77 & -0.10 (-0.13) & 0.10  (0.13) & \citet{li_2011a}\\ 
0.01 & 0.82 & -0.26 (N.A.) & 0.26  (N.A.) & \citet{cappellaro_1999}\\ 
0.03 & 0.82 & -0.32 (N.A.) & 0.32  (N.A.) & \citet{mannucci_2005}\\ 
0.025-0.050 & 0.81 & -0.24 (N.A.) & 0.33  (0.04) & \citet{dilday_2010}\\ 
0.073 & 0.71 & -0.08 (-0.06) & 0.08  (0.10) & \citet{frohmaier_2019}\\ 
0.05-0.15 & 1.60 & -0.85 (-0.58) & 1.46  (0.58) & \citet{cappellaro_2015}\\ 
0.075-0.125 & 0.76 & -0.13 (0.00) & 0.15  (0.08) & \citet{dilday_2010}\\ 
0.11 & 1.08 & -0.29 (N.A.) & 0.29  (N.A.) & \citet{strolger_2003}\\ 
0.11 & 0.72 & -0.18 (-0.09) & 0.08  (0.05) & \citet{graur_2013}\\ 
0.13 & 0.58 & -0.20 (-0.15) & 0.20  (0.15) & \citet{blanc_2004}\\ 
0.15 & 0.93 & -0.67 (-0.17) & 0.67  (0.67) & \citet{rodney_2010}\\ 
0.125-0.175 & 0.90 & -0.10 (-0.01) & 0.11  (0.10) & \citet{dilday_2010}\\ 
0.16 & 0.41 & -0.26 (-0.35) & 0.26  (0.17) & \citet{perrett_2012}\\ 
0.175-0.225 & 1.01 & -0.09 (-0.02) & 0.09  (0.24) & \citet{dilday_2010}\\ 
0.2 & 0.58 & -0.23 (N.A.) & 0.23  (N.A.) & \citet{horesh_2008}\\ 
0.25 & 1.05 & -0.76 (-1.02) & 1.75  (0.35) & \citet{rodney_2014}\\ 
0.15-0.35 & 1.14 & -0.35 (-0.29) & 0.38  (0.29) & \citet{cappellaro_2015}\\ 
0.225-0.275 & 1.06 & -0.08 (-0.03) & 0.09  (0.53) & \citet{dilday_2010}\\ 
0.26 & 0.82 & -0.20 (-0.20) & 0.20  (0.17) & \citet{perrett_2012}\\ 
0.3 & 0.99 & -0.44 (N.A.) & 0.47  (N.A.) & \citet{botticella_2008}\\ 
0.275-0.325 & 1.27 & -0.10 (-0.05) & 0.11  (1.15) & \citet{dilday_2010}\\ 
0.35 & 0.99 & -0.55 (-0.09) & 0.55  (0.55) & \citet{rodney_2010}\\ 
0.35 & 1.05 & -0.17 (-0.17) & 0.17  (0.15) & \citet{perrett_2012}\\ 
0.42 & 1.34 & -0.93 (-0.38) & 1.22  (0.29) & \citet{graur_2014}\\ 
0.44 & 0.76 & -0.39 (-0.35) & 0.67  (0.17) & \citet{okumura_2014}\\ 
0.45 & 0.90 & -0.44 (-0.12) & 0.44  (0.44) & \citet{rodney_2010}\\ 
0.45 & 1.05 & -0.17 (-0.15) & 0.17  (0.12) & \citet{perrett_2012}\\ 
0.35-0.55 & 1.52 & -0.38 (-0.47) & 0.32  (0.47) & \citet{cappellaro_2015}\\ 
0.46 & 1.40 & -0.50 (N.A.) & 0.50  (N.A.) & \citet{tonry_2003}\\ 
0.47 & 1.22 & -0.17 (-0.26) & 0.17  (0.38) & \citet{neill_2006}\\ 
0.47 & 2.33 & -0.79 (-0.76) & 1.08  (4.84) & \citet{dahlen_2008}\\ 
0.55 & 0.93 & -0.41 (-0.20) & 0.41  (0.41) & \citet{rodney_2010}\\ 
0.55 & 1.40 & -0.17 (-0.15) & 0.17  (0.12) & \citet{perrett_2012}\\ 
0.55 & 1.52 & -0.26 (N.A.) & 0.29  (N.A.) & \citet{pain_2002}\\ 
0.62 & 3.76 & -1.66 (-0.82) & 2.57  (0.79) & \citet{melinder_2012}\\ 
0.65 & 1.40 & -0.15 (-0.17) & 0.15  (0.12) & \citet{perrett_2012}\\ 
0.55-0.75 & 2.01 & -0.52 (-0.79) & 0.55  (0.79) & \citet{cappellaro_2015}\\ 
0.65 & 1.43 & -0.50 (-0.23) & 0.50  (0.50) & \citet{rodney_2010}\\ 
0.74 & 2.30 & -1.20 (N.A.) & 0.96  (N.A.) & \citet{graur_2011}\\ 
0.75 & 1.49 & -0.55 (-0.55) & 0.79  (0.67) & \citet{rodney_2014}\\ 
0.75 & 1.98 & -0.61 (-0.41) & 0.61  (0.61) & \citet{rodney_2010}\\ 
0.75 & 1.69 & -0.17 (-0.20) & 0.17  (0.15) & \citet{perrett_2012}\\ 
0.8 & 2.45 & -0.54 (-0.35) & 0.67  (0.17) & \citet{okumura_2014}\\ 
0.83 & 3.79 & -0.79 (-1.49) & 0.96  (2.13) & \citet{dahlen_2008}\\ 
0.85 & 2.27 & -0.64 (-0.47) & 0.64  (0.64) & \citet{rodney_2010}\\ 
0.85 & 1.66 & -0.15 (-0.20) & 0.15  (0.17) & \citet{perrett_2012}\\ 
0.94 & 1.31 & -0.55 (-0.17) & 0.64  (0.38) & \citet{graur_2014}\\ 
0.95 & 2.22 & -0.73 (-0.76) & 0.73  (0.73) & \citet{rodney_2010}\\ 
0.95 & 2.24 & -0.23 (-0.35) & 0.23  (0.29) & \citet{perrett_2012}\\ 
1.05 & 2.30 & -0.82 (-1.20) & 0.82  (0.82) & \citet{rodney_2010}\\ 
1.1 & 2.16 & -0.35 (-0.38) & 0.35  (0.29) & \citet{perrett_2012}\\ 
1.14 & 2.06 & -0.53 (-0.30) & 0.70  (0.30) & \citet{okumura_2014}\\ 
1.21 & 3.85 & -0.85 (-0.93) & 1.05  (1.11) & \citet{dahlen_2008}\\ 
1.23 & 2.45 & -0.82 (N.A.) & 0.73  (N.A.) & \citet{graur_2011}\\ 
1.25 & 1.87 & -0.64 (-0.67) & 0.90  (0.99) & \citet{rodney_2014}\\ 
1.59 & 1.31 & -0.64 (-0.26) & 0.99  (0.15) & \citet{graur_2014}\\ 
1.61 & 1.22 & -0.67 (-0.41) & 1.14  (0.55) & \citet{dahlen_2008}\\ 
1.69 & 2.97 & -1.08 (N.A.) & 1.57  (N.A.) & \citet{graur_2011}\\ 
1.75 & 2.10 & -0.87 (-0.82) & 1.31  (1.46) & \citet{rodney_2014}\\ 
2.25 & 1.43 & -1.11 (-0.70) & 2.77  (1.31) & \citet{rodney_2014}\\ 

\end{tabular}
\end{center}
\caption{The observations used for Type Ia comparison. Adapted from \citet{strolger_2020} with additional observations from \citet{melinder_2012} and \citet{li_2011a}, redshift ranges if available, and updated rates for \citet{madgwick_2003} from from \citet{graur_2013}. Uncertainty is split between the statistic and systematic uncertainty. If the separate numbers are available, the latter is in between brackets \label{tab:Type_Ia_obs}}
\end{table*}

\begin{table*}
\begin{center}
\begin{tabular}{c|c|c|c|c}     
\textbf{Redshift} & \textbf{Rate} & \multicolumn{2}{c}{\textbf{Uncertainty}}  & \textbf{ Reference} \\
& \multicolumn{3}{c}{[$10^5$ $h^{3}$ yr$^{-1}$ Gpc$^{-3}$]} & \\
\hline
0 & 1.81 & -0.20 (-0.44) & 0.20 (0.50) & \citet{li_2011a}\\ 
0.01 & 1.25 & -0.50 (N.A.) & 0.50 (N.A.) & \citet{cappellaro_1999}\\ 
0.028 & 2.65 & -0.37 (N.A.) & 0.45 (N.A.) & \citet{frohmaier_2021}\\ 
0.03-0.09 & 3.09 & -0.32 (-0.44) & 0.32 (0.44) & \citet{taylor_2014}\\ 
0.05-0.15 & 3.29 & -1.55 (-1.43) & 1.81 (1.43) & \citet{cappellaro_2015}\\ 
0.075 & 3.03 & -0.76 (-0.32) & 0.96 (0.12) & \citet{graur_2015}\\ 
0.1-0.5 & 8.75 & -2.74 (-1.66) & 3.73 (3.03) & \citet{dahlen_2012}\\ 
0.1-0.5 & 6.21 & -1.57 (N.A.) & 2.33 (N.A.) & \citet{strolger_2015}\\ 
0.1-0.5 & 9.59 & -5.19 (-4.23) & 8.98 (5.77) & \citet{melinder_2012}\\ 
0.15-0.35 & 3.53 & -0.79 (-1.37) & 0.79 (1.37) & \citet{cappellaro_2015}\\ 
0.21 & 3.35 & -0.99 (-1.05) & 1.25 (1.22) & \citet{botticella_2008}\\ 
0.26 & 6.41 & -2.04 (N.A.) & 2.33 (N.A.) & \citet{cappellaro_2005}\\ 
0.3 & 4.14 & -0.87 (-0.70) & 0.87 (-0.93) & \citet{bazin_2009}\\ 
0.4-0.9 & 8.16 & -5.83 (N.A.) & 13.12 (N.A.) & \citet{petrushevska_2016}\\ 
0.5-0.9 & 10.73 & -2.10 (N.A.) & 2.80 (N.A.) & \citet{strolger_2015}\\ 
0.5-0.9 & 21.55 & -4.43 (-4.66) & 5.42 (9.33) & \citet{dahlen_2012}\\ 
0.5-0.9 & 18.66 & -9.10 (-6.15) & 15.45 (10.64) & \citet{melinder_2012}\\ 
0.5-1.0 & 20.12 & -15.74 (N.A.) & 28.86 (N.A.) & \citet{graur_2011}\\ 
0.9-1.3 & 8.95 & -1.92 (N.A.) & 3.09 (N.A.) & \citet{strolger_2015}\\ 
0.9-1.3 & 27.90 & -8.16 (-8.16) & 10.96 (14.46) & \citet{dahlen_2012}\\ 
0.9-1.4 & 30.03 & -17.78 (N.A.) & 33.53 (N.A.) & \citet{petrushevska_2016}\\ 
1.3-1.7 & 9.48 & -3.85 (N.A.) & 5.92 (N.A.) s& \citet{strolger_2015}\\ 
1.4-1.0 & 31.49 & -25.95 (N.A.) & 71.14 (N.A.) & \citet{petrushevska_2016}\\ 
1.7-2.1 & 9.21 & -5.16 (N.A.) & 9.83 (N.A.) & \citet{strolger_2015}\\ 
2.1-2.5 & 17.99 & -10.26 (N.A.) & 19.71 (N.A.) & \citet{strolger_2015}\\ 

\end{tabular}
\end{center}
\caption{The observed CCSN rates from literature. Uncertainty is split between the statistic and systematic uncertainty. If the separate numbers are available, the latter is in between brackets.\label{tab:CCSN_obs}}
\end{table*}

\begin{table*}
\begin{center}
\begin{tabular}{c|c|c|c|c}     
\textbf{Redshift} & \textbf{Rate} & \multicolumn{2}{c}{\textbf{Uncertainty}}  & \textbf{ Reference} \\
& \multicolumn{3}{c}{[$h^{3}$ yr$^{-1}$ Gpc$^{-3}$]} & \\
\hline
0.17 & 102 & -38 & 73 & \citet{frohmaier_2021}\\ 
0.17 & 89 & -73 & 215 & \citet{quimby_2013}\\ 
0-1.6 & 117 & N.A. & N.A. & \citet{zhao_2020}\\ 
1.13 & 265 & -105 & 222 & \citet{prajs_2017}\\ 
2.0-4.0 & 1118 & -559 & 559 & \citet{cooke_2012}\\ 
2.5-3.5 & 1166 & -1166 & 1166 & \citet{moriya_2019}\\ 

\end{tabular}
\end{center}
\caption{The observed SLSN Type I rates from literature with the combined statistical and systematic uncertainties are given. Adapted from \citet{frohmaier_2021}. \label{tab:SLSN_obs}}
\end{table*}

\begin{table*}
\begin{center}
\begin{tabular}{c|c|c|c|c}    
\textbf{Redshift} & \textbf{Rate} & \multicolumn{2}{c}{\textbf{Uncertainty}}  & \textbf{ Reference} \\
& \multicolumn{3}{c}{[$10^5$ $h^{3}$ yr$^{-1}$ Gpc$^{-3}$]} & \\
\hline
0 & 1.15 & -0.36 (N.A.) & 0.36 (N.A.) & \citet{li_2011a}\\ 
0.15-0.35 & 2.01 & -0.52 (-0.70) & 0.47 (0.70) & \citet{cappellaro_2015}\\ 

\end{tabular}
\end{center}
\caption{The observed Type II rates from literature. Uncertainty is split between the statistic and systematic uncertainty. If the separate numbers are available, the latter is in between brackets. \label{tab:Type_II_obs}}
\end{table*}


\bsp	
\label{lastpage}
\end{document}